\documentclass[aps,superscriptaddress,twocolumn,showkeys,amsmath,amssymb,pre,showpacs,floatfix]{revtex4-1} % {{{
\usepackage[utf8]{inputenc} 
\usepackage{amsmath}
\usepackage{graphicx}
\usepackage{bm}
\usepackage{color}
\usepackage{float}

\newcommand{\rmc}{\mathrm{c}}
\newcommand{\rmd}{\mathrm{d}}
\newcommand{\rme}{\mathrm{e}}
\newcommand{\rmi}{\mathrm{i}}
\providecommand{\openone}{\leavevmode\hbox{\small1\kern-3.8pt\normalsize1}}
\newcommand{\ie}{\textit{i.e.} }

\newcommand{\mcN}{\mathcal{N}}
\newcommand{\tr}{\mathrm{tr}}
\newcommand{\tre}{\tr_\rme}

\newcommand{\One}{\openone}
\newcommand{\eps}{\varepsilon}

\newcommand{\la}{\langle}
\newcommand{\ra}{\rangle}
\newcommand{\lla}{\left\langle}
\newcommand{\rra}{\right\rangle}

\newcommand{\eref}[1]{Eq.~(\ref{#1})}

\newcommand{\Eref}[1]{Eq.~(\ref{#1})}
\newcommand{\Sref}[1]{Sec.~\ref{#1}}
\newcommand{\Fref}[1]{Fig.~\ref{#1}}

\graphicspath{{./figs/}}
\newcommand{\mt}[4]{\begin{pmatrix}
                    #1 & #2 \\[5pt]
                    #3 & #4
                 \end{pmatrix}}

\begin{document}
\title{Markovian and non-Markovian dynamics induced by a generic environment}
\author{M. Carrera}
\affiliation{Instituto de Física, Benemérita Universidad Aut\' onoma de
Puebla, Apartado Postal J-48, Puebla 72570, M\' exico.}
\author{T. Gorin}
\affiliation{Departamento de F\'\i sica, Universidad de Guadalajara,
  Blvd. Marcelino Garc\'\i a Barragan y Calzada Ol\'\i mpica, Guadalajara
  C.P. 44840, Jal\'\i sco, M\' exico.}
\author{C. Pineda}
\affiliation{Instituto de F\'\i sica, Universidad Nacional Aut\' onoma de 
  M\' exico, M\' exico D.F. 01000, M\' exico}

\begin{abstract}
We study the open dynamics of a quantum two-level system coupled to an 
environment modeled by random matrices. Using the quantum channel 
formalism, we investigate different quantum Markovianity measures and criteria.
A thorough analysis of the whole parameter space, reveals a wide range of 
different regimes, ranging from strongly non-Markovian to Markovian dynamics. 
In contrast to analytical models, all non-Markovianity measures and criteria 
have to be applied to data with fluctuations and statistical uncertainties. 
We discuss the practical usefulness of the different approaches.
\end{abstract}


\maketitle

% }}}
\section{\label{I} Introduction} % {{{
Open quantum systems have been of interest for a long time~\cite{landau}. The
interest stems from the natural separation of a quantum system into 
a central system of interest, and an
uninteresting or uncontrollable part, usually denoted by {\it environment}.
In 1967, Lindblad~\cite{Lindblad1976}, Gorini, Kossakowski and
Sudarshan~\cite{Gorini1976} arrived at the so-called
Lindblad master equation to describe the evolution of a central system weakly 
interacting with a memoryless environment.
This equation has been of paramount importance in
the field as one can, both, analytically solve several instances of the
equation~\cite{Prosen2008}, and  describe accurately a wide range of
experimental situations~\cite{carmichael1999statistical}. The dynamics produced
by such an equation is called {\it quantum Markovian dynamics}.
Recently there has been an effort to classify and understand systematically
open quantum systems which lie outside this description. 

Definitions and measures for quantum non-Markovianity (NM) have 
received considerable interest in the last 10 years or 
so~\cite{PhysRevLett.103.210401,Rivas2010}. They are meant to characterize 
quantum processes (viz. the dynamics of open quantum systems) which cannot 
be described by a master equation with constant Lindblad operators.
For such systems, there might be hierarchy of stronger/weaker non-Markovianity 
in the sense that it is easy/not-easy to describe the quantum channel by some 
effective evolution equation within the systems state space alone. A second 
quite different idea is that NM is a
feature which might be taken advantage of in order to perform certain tasks.
Both are valid points of view, but it is not clear to what extend some
of the popular definitions and measures of NM provide relevant information
with respect to these questions. Several reviews of the field can be found in
Refs.~\cite{Breuer2012,RHP14,RevModPhys.88.021002,RevModPhys.89.015001}.

Non-Markovianity has been studied extensively, for quantum processes 
with analytical solution, implying no fluctuations or uncertainties, 
applications to more realistic processes are relatively rare~\cite{Liu2013,
PhysRevA.98.053862}.

In the present paper, we study a quantum channel derived from the coupling of 
a two-level system to a ``generic'' quantum environment. We use random matrix 
theory (RMT) to describe that environment. Choosing the Hamiltonian in the 
environment from the Gaussian unitary ensemble (GUE), we find a variety of 
different behaviors in the relevant parameter space, ranging from Markovian 
(Lindblad-Dynamics) to strongly NM behavior. This model is ideal to discuss the 
questions raised above. Since the model has no known analytical solution, all 
criteria and measures must be calculated numerically, with unavoidable 
statistical errors, which resembles an experimental situation in which finite 
statistics come into play.  

In \Sref{G} we present the model and find the structure of the quantum channel
describing the dynamics of the system. In \Sref{N}, we introduce the NM
measures, which we will use in \Sref{S} for analysis. In~\Sref{S} we compare
and interpret the different measures for the whole available region in the 
parameter space. We finish with some closing remarks in \Sref{C}.
% }}}

\section{\label{G} Model} % {{{
The model to be used has been introduced in Ref.~\cite{CGS14}, which focussed 
on the derivation of an analytical description in the linear response regime. 
We find it suitable for the purpose of the present work, since it is a generic 
model, which, nontheless shows a broad range of different behaviors with 
respect to quantum NM. In this section, we describe the Hamiltonian of our 
system, and the quantum channel formalism used for a complete description of 
its dynamics. 

\subsection{The Hamiltonian} % {{{ 
We consider a two level system (qubit) coupled to an environment. The Hilbert 
space of the qubit will be labeled by the subindex $_\rmc$ and that of the 
environment by the subindex $_\rme$. We assume that the dynamics in the whole 
Hilbert space is unitary, with the evolution governed by the Hamiltonian
\begin{equation}
H_\lambda =  
\frac{\Delta}{2} \sigma_z \otimes \One_\rme  + \One_\rmc \otimes  H_\rme + 
\lambda\; v_\rmc \otimes V_\rme \; .
\label{G:Hlam} 
\end{equation}
All subindices of operators in the right hand side indicate the subspace in 
which they act, except for $\sigma_{z}$, which is a Pauli matrix acting on the 
qubit. $\Delta$ is the level splitting in the qubit and the parameter $\lambda$ 
controls the strength of the coupling between the central system and the 
environment. The first two terms in \eref{G:Hlam} represent the free evolution 
of both central system and environment, while the third term provides the 
coupling which is assumed to be separable. The Hamiltonian of the environment 
$H_\rme$ shall be chosen from the Gaussian unitary ensemble to provide 
generality to the results discussed here~\cite{Meh2004}. We measure time in 
units of the Heisenberg time of $H_\rme$ and energy in units of the means 
level spacing in the center of the spectrum of $H_\rme$; in such units, 
$\hbar = 1$.  The density matrix of the central system for a time $t$ is 
given by
\begin{equation}
\varrho_\rmc(t) = \tre\left[ \lla\, \rme^{-\rmi H_\lambda t}\, 
   \varrho_\rmc \otimes \varrho_\rme\, \rme^{\rmi H_\lambda t}\,\rra \right],
\label{G:reducedDyn}\end{equation}
where we have chosen a product state as the initial state of central system and
environment. Just as $H_\rme$, $V_\rme$ is also chosen from the Gaussian
unitary ensemble, and the
angular brackets denote an ensemble average over both random matrices. The 
magnitude of the  matrix elements $[V_\rme]_{ij}$ is chosen such that 
$\la [V_\rme]_{ij} [V_\rme]_{kl}\ra = \delta_{il}\delta_{jk}$.

The most general form of $v_\rmc$ has a parallel and perpendicular component, 
with respect to the internal Hamiltonian $\sigma_z$. If there is only a 
parallel component ($v_\rmc \propto\sigma_z $) the qubit dynamics becomes 
dephasing. The channel acting on the qubit can be obtained in terms of the
fidelity amplitude for the Hamiltonians $H_\rme \pm \lambda V_\rme$.
This is already a very rich case; however, it has been considered 
before~\cite{GCZ97,GPSS04}. The other limiting case is when the coupling
is perpendicular to the internal Hamiltonian, which
is the case we are studying here; see also~\cite{CGS14}. Thus we simply set,
without loosing any generality,
\begin{equation}
v_\rmc = \sigma_x \; ,
\label{G:vcissigx}\end{equation}
where $\sigma_x$ is one of the three Pauli matrices.
% }}}

\subsection{Quantum channel formalism} %{{{
We describe the reduced dynamics of the qubit with the quantum channel 
formalism, which means that the evolution of the system state is described in
terms of a linear time dependent map $\Lambda_t$ acting on the space of
density matrices $\mathcal{S}(\mathbb{C}^2)$ of the central system. The map 
$\Lambda_t$ takes an arbitrary initial state, and returns the state evolved 
according to \Eref{G:reducedDyn} for a time $t$,
\begin{equation}
\Lambda_t \; :\; \varrho_\rmc \to \varrho_\rmc(t)= \Lambda_t[\varrho_\rmc] \; .
\label{G:QuChan}\end{equation}
Since the image of this map is a density matrix, $\Lambda_t$ has two 
properties: (i) it preserves the trace of the argument and (ii) is completely 
positive. Maps with these characteristics will be referred to as 
\textit{quantum channels}. We will also be interested in more general linear 
operators that preserve the trace but, even though map hermitian operators to
hermitian operators, are not necessarily completely positive. In fact, we
shall consider maps that are generally non positive.  We will call such maps 
\textit{quantum maps}. In this language, a quantum channel is a quantum map, 
but not necessarily the other way around. 

A quantum map $K$, and any linear map,  is determined by its action on a basis;
consider the computational basis $\{ |a\ra\la b|\ \}_{a,b\in \{0, 1\}}$ and
arrange the resulting elements in the matrix
\begin{equation}
 C_K = \begin{pmatrix} 
      K[\, |0\ra\la 0|\, ] & K[\, |0\ra\la 1|\, ] \\
      K[\, |1\ra\la 0|\, ] & K[\, |1\ra\la 1|\, ] 
      \end{pmatrix} \; .
\label{IntroChoiMatrix}
\end{equation}
This is the so called Choi-matrix representation~\cite{Choi1975285,zimanbook2012}. Since $K$ maps hermitian matrices to hermitian matrices, $C_K$ must also be
hermitian. %\par

It can be seen that the Choi matrix can be obtained by
applying the extended map ${\rm id} \otimes K$ to a Bell state in the Hilbert
space of two qubits:
\begin{equation}
 C_K = 2\, ({\rm id} \otimes K)\, [\, |\text{Bell}\ra\la \text{Bell}|\, ] \; , 
\end{equation}
where $|\text{Bell}\ra= (|00\ra + |11\ra/\sqrt{2}$.
Thus, if $K$ is a quantum channel, $1/2\, C_K$ is a two qubit
density matrix, which in turn implies that $C_K \ge 0$, i.e.  all of its
eigenvalues must be positive or equal to zero. 

We will now outline the procedure to construct the Choi matrix, based 
on the dynamics. The procedure will rely on two properties of our
particular channel. First, it is unital, which means that the identity is 
mapped onto the identity. Second, the evolution of a diagonal 
operator remains diagonal, i.e. if $\varrho_\rmc$ is diagonal, so is 
$\varrho_\rmc(t)$ in \eref{G:QuChan}. The proof of both properties is found in 
the Appendix \ref{app:DS}. Let $\rho_\rmc^{x,y,z}$ be the density matrices 
associated with the $+1$ eigenvalues of the corresponding Pauli matrices. 
From the second property and trace conservation, we find that
\begin{equation}
\Lambda_t[ \rho_\rmc^{z} ] =
\begin{pmatrix} 
r & 0\\ 
0 & 1 - r 
\end{pmatrix}
\end{equation}
and from the first property, plus linearity of quantum maps, we get
\begin{equation}
\Lambda_t[\rho_\rmc^{x}\, ] = \frac{1}{2}
\begin{pmatrix} 
   1 & z_x^*\\
   z_x & 1
\end{pmatrix}\; ,\quad
\Lambda_t[\rho_\rmc^{y}\, ]
 = \frac{1}{2}\begin{pmatrix} 
   1 & z_y^*\\
   z_y & 1
\end{pmatrix}\; 
\end{equation}
for some time dependent functions $r$, $z_x$ and $z_y$. From these equations, 
one can see that the dynamics in the $z$ axis is decoupled from the ones in 
the $xy$ plane. If we define $ z_{1,2}= (z_x \mp \rmi z_y)/2$, we can write 
directly
\begin{equation}
C_t \equiv C_{\Lambda_t} = \begin{pmatrix} r & 0 & 0 & z_1^*\\
         0 & 1-r & z_2 & 0\\
         0 & z_2^* & 1-r & 0\\
         z_1 & 0 & 0 & r\end{pmatrix} \; .
\label{G:ChoiLam}\end{equation}
Notice that  $r|_{t=0}= z_1|_{t=0}= 1$, and $z_2|_{t=0}= 0$.
% }}}
% }}}

\section{\label{N} Non-Markovianity measures} % {{{
We consider NM as a property of a quantum process $\Lambda_t$, which is a 
one-parameter family of quantum channels with $t\in \mathbb{R}_0^+$ and
$\Lambda_0 = {\rm id}$. The NM criteria and measures used here are based on 
two different concepts, (i) divisibility and (ii) contractivity. Both of
them require knowledge of the intermediate quantum map
\begin{equation}
\Lambda_{t+\eps,t} = \Lambda_{t+\eps} \circ \Lambda_t^{-1} \quad :\quad
  \varrho \to \Lambda_{t+\eps}\big [\, \Lambda_t^{-1}[\varrho]\, \big ] \; .
\label{N:relqumap}\end{equation}
In Appendix~\ref{aR}, we calculate the Choi representation of this 
intermediate quantum map with the following result:
\begin{equation}
C_{t+\eps,t} =\begin{pmatrix} 
        q & 0 & 0 & Z_1^*\\
         0 & 1-q & Z_2 & 0\\
         0 & Z_2^* & 1-q & 0\\
         Z_1 & 0 & 0 & q
         \end{pmatrix}\; , 
\label{N:ChoiRelLam}
\end{equation}
with $D=|z_1|^2-|z_2|^2$, $ Z_1= (z_1'z_1^* - z_2' z_2^*)/D$, 
$Z_2= (z_2' z_1  - z_1' z_2)/D$ and $q= (r+r'-1)/(2r-1)$. The parameters $r'$, 
$z_1'$ and $z_2'$ are the same as $r$, $z_1$ and $z_2$ but calculated at a 
time $t+\eps$. When $D=0$ or $2r-1 = 0$, $\Lambda_t$ is not invertible, and 
therefore $\Lambda_{t+\eps,t}$ may not exist. 
% }}}

\subsection{\label{subsec:divi}Divisibility}  % {{{
A quantum process $\Lambda_t$ is divisible if and only if for any $t,\eps > 0$
it holds that $\Lambda_{t+\eps}$ can be written as the composition
$\Lambda_{t+\eps} = \Lambda_x \circ \Lambda_t$, with $\Lambda_x$ 
being a valid quantum channel. 
Here, $\Lambda_x$ can be
identified with the intermediate quantum map $\Lambda_{t+\eps,t}$ given in 
Eq.~(\ref{N:relqumap}). Hence the divisibility of a quantum process is 
equivalent to all intermediate quantum maps being valid quantum channels. 
Formally speaking, the quantum process $\Lambda_t$ is divisible if and only if
\begin{itemize}
\item[(a)] $\Lambda_t$ is invertible for almost all $t\in \mathbb{R}_0^+$, and
\item[(b)] $\forall t,\eps > 0 \; :\; 
   \Lambda_{t+\eps,t} = \Lambda_{t+\eps} \circ \Lambda_t^{-1}$ is a valid
   quantum channel if it exists.
\end{itemize}
In condition (a), we allow $\Lambda_t$ to be non-invertible at a finite 
(countable) number of points in time. 
In condition (b), we check the complete positivity of the 
intermediate map only for those $t$, where $\Lambda_t$ is invertible.
% }}}

\paragraph{RHP-Markovianity} % {{{
One of the definitions of NM is given in terms of the divisibility 
of the quantum process under consideration. Following Rivas 
et al.~\cite{RiHuPl10}, we call a quantum process $\Lambda_t$, RHP-Markovian 
if and only if the two conditions above are fulfilled. 

To check for the complete positivity of the intermediate quantum map, Rivas 
et al. consider the trace norm of the associated Choi matrix defined as
$\|C_{t+\eps,t}\|_1 = \sum_j |\lambda_j|$, where $ \lambda_j$ are the 
eigenvalues of $C_{t+\eps,t}$. Since the sum of the eigenvalues always is 
equal to the dimension of the Hilbert space of the physical system, due to 
trace preservation, any negative eigenvalue will necessarily lead to 
$\|C_{t+\eps,t}\|_1$ being larger than two. In addition, since the composition 
of two quantum channels is again a quantum channel, it is sufficient to check 
complete positivity for infinitesimal $\eps$, only. Hence, Rivas et al. 
define the function 
\begin{equation}
g(t) = \lim_{\eps\to 0} \frac{1}{2\eps}\, 
   \big (\,  \|C_{t+\eps,t}\|_1 - 2\, \big )\; ,
\label{N:gdefgeneral}\end{equation}
which is zero if the intermediate map is completely positive, and 
greater than zero otherwise. Moreover, to define a measure of the degree of
non-Markovianity of the process, the authors integrate this function over time. 
We shall label this quantity as
\begin{equation}
\mathcal{N}_{\rm RHP}(t)= \int_0^t\rmd\tau\; g(\tau) \; .
\label{eq:IRPLdef}\end{equation}
Notice that $\mathcal{N}_{\rm RPL}(t)=0$ if and only if
the process is divisible for all times up to $t$. 
% }}}

\paragraph{Application to our model} % {{{
In order to check the RHP-Markovianity (divisibility) of $\Lambda_t$ as defined
in Sec.~\ref{G}, we use the Choi representation, \eref{N:ChoiRelLam} of 
the intermediate quantum map $\Lambda_{t+\eps,t}$. 
The eigenvalues of $C_{t+\eps,t}$ are 
\begin{equation}
\lambda_{1,2}=q\pm|Z_1|,\;\; \lambda_{3,4}=(1-q)\pm|Z_2|.
\end{equation}
Hence, the eigenvalues are non-negative
if and only if (i) $|Z_1| \le q$ and (ii) $|Z_2| \le 1-q$. 

Since we can limit to infinitesimal $\eps$, we expand the different
functions in \eref{N:ChoiRelLam} around $t$ and obtain, to first order,
\begin{align}
Z_1 &= 1+ \frac{\eps}{D}(\dot{z}_1z_1^*-\dot{z}_2z_2^*) \; ,\\
Z_2 &= \frac{\eps}{D}(\dot{z}_2z_1-\dot{z}_1z_2) \notag\\
q &
=1+\frac{\eps\dot{r}}{2r-1} \; ,
\end{align}
where we have used the fact that $r'= r + \eps\, \dot{r}$, 
$z_1'= z_1 + \eps\, \dot{z}_1$, and $z_2'= z_2 + \eps\, \dot{z}_2$.

The two conditions (i) and (ii) can now be written as
\begin{align}
  ({\rm i})&\qquad 1 - \eps\, \delta_1 \le 1 - \eps\, \delta_q \; , \
  \quad\Leftrightarrow\quad 
  \delta_1 \ge \delta_q \notag\\
({\rm ii})&\qquad |Z_2| \le 1 - q \quad\qquad\quad\Leftrightarrow\quad
\delta_2 \le \delta_q \; ,
\end{align}
where we introduced
\begin{align}
 \delta_1 &= - \frac{1}{D}\, {\rm Re}[ \dot{z}_1 z_1^* - \dot{z}_2 z_2^* ]\;,\\
\delta_q &= \frac{-\, \dot{r}}{2r-1} \ge 0 \;,\\
\intertext{and}\quad \delta_2 &=  \frac{|\dot{z}_2z_1-\dot{z}_1z_2|}{|D|} \; . 
\label{def:deltas}\end{align}
Additionally we used the fact that for any complex number $c$, $|1+\eps c| 
= 1 + \eps c + \mathcal{O}(\eps^2)$ on the expression for $|Z_1|$.
Finally, we combine the two inequalities into
\begin{equation}
\delta_2 \le \delta_q \le \delta_1 \; .
\label{divcondition}
\end{equation}

We can now relate our inequalities to the criterium of Rivas et al. as follows.
In our case, the trace norm of the Choi matrix, can be written as
\begin{align}
&||C_{t+\eps,t}||_{1} = \big |q + |Z_1|\, \big | + \big |q - |Z_1|\, \big | 
   + \big |1-q + |Z_2|\, \big | \notag\\
&\qquad\qquad\qquad + \big |1-q - |Z_2|\, \big | \notag\\
&= 2 - \eps\, \big [\, \delta_q + \delta_1 + |\delta_1 - \delta_q|
  + |\delta_q + \delta_2| + |\delta_q - \delta_2|\, \big ]\; .
\end{align}
This yields 
\begin{align}
g(t) = \frac{|\delta_1 - \delta_q| + |\delta_q + \delta_2| 
         + |\delta_q - \delta_2| -\delta_q -\delta_1}{2} \; .
\label{eq:goft}\end{align}

We showed that non-negativity of the eigenvalues is equivalent
to the double inequality $\delta_2 \le \delta_q \le \delta_1$. Then we saw that
it is also equivalent to $g(t) = 0$. This means that the double inequality 
holds if and only if $g(t) = 0$.
% }}}

\subsection{\label{NC} Contractivity}  % {{{
Markovianity of classical stochastic processes imply that 
probabilities distributions
decrease their Kolmogorov distance with time~\cite{vKampen07}. This 
is interpreted as a loss of information of the initial conditions.
Carrying this ideas to a quantum level results in a definition 
of Markovianity~\cite{BrLaPi09}.
Let $\rho_{1,2}(t)$ denote the evolution of two states $\rho_{1,2}$. 
We define
\begin{equation}
\sigma(\rho_{1}, \rho_2, t)
=\frac{\rmd }{\rmd t}T[\varrho_1(t),\varrho_2(t)]
\label{eq:sigmaoft}
\end{equation}
where $T[\varrho_1(t),\varrho_2(t)]=\tr(|\varrho_1(t)-\varrho_2(t)|)/2$ is the
trace distance, which is directly related with the probability of distinguish
the state $\varrho_1(t)$ from the state $\varrho_2(t)$, i. e., it is their
distinguishability~\cite{NieChu00},
and $|A|=\sqrt{AA^{\dagger}}$. 
In other words, $\sigma$ is the derivative of the distance between 
the evolved states. 
We say that a process is contractive if for all $\rho_{1,2}$ and all
$t \ge 0$, we have that $\sigma(\rho_{1}, \rho_2, t) \le 0$.
A process is said to be non-Markovian if it is not contractive. 
Breuer~et al. then define the following quantity as a measure for the degree
of non-Markovianity:
\begin{equation}
\mcN_{\rm BLP}(t)=\max_{\rho_1, \rho_2} 
\int_{0\le \tau \le t ,\sigma > 0} \rmd \tau\;\;
\sigma(\rho_{1}, \rho_2, \tau) \;.
\label{eq:IBLPdef}
\end{equation}

The calculation of this measure is greatly simplified when the process 
acts on a qubit. To perform
this maximization, one should consider only pure, orthogonal initial 
states~\cite{Wis12}. Indeed, we found that $T[\varrho_1(t),\varrho_2(t)]$ only 
depends on the vector 
difference between the representations of initial states in the Bloch ball 
(see Appendix~\ref{app:sec:tracedist}). Moreover, the 
distance between the two points representing the initial states enters
as a homogeneous scale factor. It therefore possible, restricting the 
maximum search to such cases, where $\rho_1$ is a pure state, and
$\rho_2$ the uniform mixture. If the pure state $\rho_1$ is 
parametrized in spherical coordinates by the angles $\theta$ and $\phi$,
we obtain [cf. Eq.~(\ref{eq:app:tdqbgeneral})]
\begin{align}
&\sigma(\rho_1,\One/2,t) =  \frac{1}{2}\, \frac{\rmd}{\rmd t}\, 
   ||\sin\theta(\cos\phi\sigma_x+\sin\phi\sigma_y)+ \cos\theta\sigma_z||
\notag\\
&\qquad = \frac{1}{2}\, \frac{\rmd}{\rmd t}\,
   \sqrt{(2r-1)^2\cos^2\theta+ M(\phi)\sin^2\theta} \; ,
\label{eq:lainemeasure1}\end{align}
with $M(\phi)=|z_1+z_2\, \rme^{-2\rmi\phi}|^2$. 

In what follows, we derive a criteria for 
NM based on the contractivity, which can be compared to the criteria
Eq.~(\ref{divcondition}) obtained in Sec.~\ref{subsec:divi}, based on 
divisibility. Since $\theta$ may be chosen freely in 
Eq.~(\ref{eq:lainemeasure1}),
a given process is Markovian in the sense of Breuer et al., if and only if 
both functions, $(2r-1)$ and $M(\phi)$ are non-increasing at all times. 
In other words if
\begin{align}
\frac{{\rm d}}{{\rm d} t}(2r-1)^2&\leq0\label{condidista}\\
\frac{{\rm d}}{{\rm d} t}M(\phi)&\leq0\label{condidistb}
\end{align}
for all times. Condition \eqref{condidista} becomes $-\dot{r}(2r-1)=
\delta_q(2r-1)^2\geq0$ which in turn is equivalent to $\delta_q\geq0$
(at least as long as $2r - 1 \neq0$, \ie away from the points
where $\Lambda_t$ is not invertible). To consider 
condition~\eqref{condidistb}, we expand $M(\phi)$ as
\begin{equation}
M(\phi)=A+\cos(2\phi)\;B-\sin(2\phi)\;C,
\end{equation}
where $A=|z_1|^2+|z_2|^2$, $B=2\,{\rm Re}(z_1z_2^*)$ and 
$C=2\, {\rm Im}(z_1z_2^*)$. 
Setting $\dot{B}=R\cos\alpha$ and $\dot{C}=R\sin\alpha$,
we may write 
\begin{align}
\frac{\rmd}{\rmd t}M(\phi)&=
\dot{A}+\sqrt{\dot{B}^2+\dot{C}^2}\cos(2\phi+\alpha).
\end{align}
From this, it is clear that the largest time derivative of $M(\phi)$ is
$\dot M_{\max}=\dot{A}+\sqrt{\dot{B}^2+\dot{C}^2}$.
With $Z=z_1z_2^*$, we find $\dot{B}=\dot{Z}+\dot{Z}^*$ and
$\dot{C}=-\rmi\;(\dot{Z}-\dot{Z}^*)$ such that
\begin{equation}
\dot M_{\rm max}=
\dot{A}+\sqrt{(\dot{Z}+\dot{Z}^*)^2- (\dot{Z}-\dot{Z}^*)^2}
=\dot{A}+2|\dot{Z}|.
\label{def:sigmamax}\end{equation}
To summarize, for the process to be Markovian (in the sense of
contractivity), it is required that both,
$\delta_q \ge 0$ and $\dot M_{\max}\leq0$. This is equivalent to
\begin{equation}
\label{condidistc}
 \delta_q \ge 0 \; , \qquad 
 \delta^{\rm C}_1 \ge \delta^{\rm C}_2 \; ,
\end{equation}
where
\begin{equation}
\delta^{\rm C}_2 = |\dot z_1 z_2^* + z_1 \dot z_2^*|\; , \quad
\delta^{\rm C}_1 = -{\rm Re}(\dot z_1 z_1^* + \dot z_2 z_2^*) \; . 
\end{equation}
This double inequality is the analog of Eq.~(\ref{divcondition}), which has
been derived as criterium for Markovianity in the sense of divisibility. 

For unital maps like the one considered here, contractivity of the trace
distance can be identified with positivity. That means that our quantum
process $\Lambda_t$ is contractive if and only if all intermediate maps are 
positive. Note that divisibility is defined as all intermediate maps being 
completely positive. Thus, there may be processes which are contractive but not
divisible~\cite{Chr11}. Therefore, we may find regions in the parameter space
of our system, where the dynamics is contractive (i.e. BLP-Markovian) but not 
divisible (i.e. RHP-Markovian). A recent, comprehensive discussion on different
criteria for divisible and contractive processes can be found in 
Ref.~\cite{MoDaGo19}.
% }}}

\subsection{Maximal recovery} % {{{
This quantifier of non-Markovianity can be based on any capacity-like property
of the channel.  In fact, we use distinguishability, maximized over states,
just as the BLP-measure. 
However, instead of summing up all the small increments, we 
search for the maximum distinguishability recovery over the whole quantum 
process~\cite{PGDWG16}:
\begin{multline}
\mathcal{N}_{\rm MDR}(t) = 
\max_{\substack{t\ge t_1\ge t_2 \ge 0 \\ \varrho_1,\varrho_2} }
\{\, T(\varrho_1(t_1), \varrho_2(t_1))  \\
   - T(\varrho_1(t_2), \varrho_2(t_2))\, \}\; ,
\label{def:MRtr}\end{multline}
where $T(\varrho_1,\varrho_2)$ is defined in the text below 
Eq.~(\ref{eq:sigmaoft}). 
The BLP-measure has been related to the backflow of information,
which can be quantified indeed in terms of the recovery of distinguishability.
However, in order to quantify the amount of information recovered, it is much
more sensible to use $\mathcal{N}_{\rm MDR}$, than integrating over all 
backflow in a process, where information is fluctuating back and forth between
system and environment. Of course there is a price to pay. It is quite more 
expensive to compute $\mathcal{N}_{\rm MDR}$, than it is to compute 
$\mathcal{N}_{\rm BLP}$, because of the additional degrees of 
freedom, $t_1$ and $t_2$.
% }}}

\section{\label{S} Numerical simulations} 
% Intro {{{
In this section, we apply different methods to characterize the 
(non-)Markovian dynamics of a generic open quantum system. 
The quantum process to be studied is obtained by numerical simulations of the 
Hamiltonian in \Eref{G:Hlam} with the initial state of the environment taken
as the maximally mixed state. This includes a Monte-Carlo sampling of a random
matrix ensemble. Therefore, the numerical data are contaminated 
by residual statistical fluctuations due to the finite size of the sample. We 
stress that residual fluctuations would be present, in 
experimental situations, also.  The dimension of the Hilbert space of the 
environment is set to $N= 200$ and the sample size is fixed to 
$N_{\rm sam}= 2400$ unless otherwise stated.

In our model system, defined in \Eref{G:Hlam}, we can 
identify three different energy scales. The average level spacing in $H_\rme$ 
(which is set equal to one), the spacing $\Delta$ between the two levels of the 
qubit, and the coupling strength $\lambda$ between qubit and environment. 
In an effort to explore the properties of our model as thoroughly as possible, 
we consider a rectangular region in the parameter space $(\Delta,\lambda)$. For
$\Delta$, the interesting range reaches from values much
smaller than the mean level spacing to values much larger, thus we choose the
limits $0.016 \le \Delta \le 16$, covering a range of three orders of 
magnitude. For 
$\lambda$, we choose the lower limit where $\lambda$ is much smaller than
the average level spacing in $H_\rme$, such that perturbation theory would be
applicable for the Hamiltonian $H_\rme + \lambda V$. The upper limit, by 
contrast, is dictated by the requirement that our model reproduces the behavior 
of the random matrix model in the limit $N\to\infty$, in such a way that finite 
size effects are still negligible. For that to hold, the spreading of 
eigenstates of $H_\rme + \lambda V$ expanded in the basis of $H_\rme$ must be 
small compared to the spectral range of $H_\rme$. Both considerations lead us
to the limits $1/32 \le \lambda \le 1/2$. A similar parameter 
range has been explored in Ref.~\cite{CGS14}.

In order to obtain reliable values for the two measures, we establish a finite
ending time for the processes to be
studied. At that time, the system will have relaxed
so much that the remaining dynamics is unusable for any practical purposes.
Our approach for finding a sensible definition for the ending time
$t_{\rm End}$ is described in Sec.~\ref{SE}. 
In Sec.~\ref{SM} we present and discuss the results for three non-Markovianity
measures. Moreover we select several representative parameter 
sets that are analyzed in detail in Sec.~\ref{SS}. In particular, we study the
dependence of the measures on the number of samples considered.
The motivation is to
understand the statistical significance of the results presented in \Sref{SM}.
At the end, this could be useful for identifying
a NM measure which is accurate, robust and significant. 
We finish this section by analyzing the local (in time) criteria for
non-Markovianity in \Sref{SC}.  We consider divisibility and contractivity, via
the corresponding inequalities \Eref{divcondition} and \Eref{condidistc}.
Both conditions are tested for the infinitesimal intermediate quantum map
$\Lambda_{t+\eps,t}$ as defined in \eref{N:relqumap}, and examine carefully 
the usefulness of such kind of expression under statistical fluctuations.  
% }}}

\subsection{\label{SE} Process ending time} % {{{
For the NM measures to be considered below (Sec.~\ref{SM}) it is essential to 
define an ending time $t_{\rm End}$ for the quantum process in question. 
However, two conflicting requirements arise. On the one hand
the ending time should be sufficiently large, such that the 
dynamics of the process is completely contained,  but on the other hand it
should also be sufficiently short, such that the contaminating contribution 
from residual statistical fluctuations remain small. 

The quantum process studied here, has the convenient property that if one 
chooses as initial state an
eigenstate of $\sigma_y$, the system converges to the uniform
mixture in the limit of long times.
Along this process, the purity $P(t)= {\rm tr}[\varrho_\rmc(t)^2]$ decays from
$P(0) = 1$ to $P(\infty) = 1/2$. 
We choose the ending time for the process at that time, where the purity 
of $\Lambda_t[\rho_\rmc^{\rm y}]$ is equal to $0.51$, which means that the 
purity has decayed to $2\%$ above its minimum value. Of course other values of
the same order are equally possible, but they do 
not change the general results of our study.

\begin{figure}%{{{
\centering
\includegraphics[width=\columnwidth]{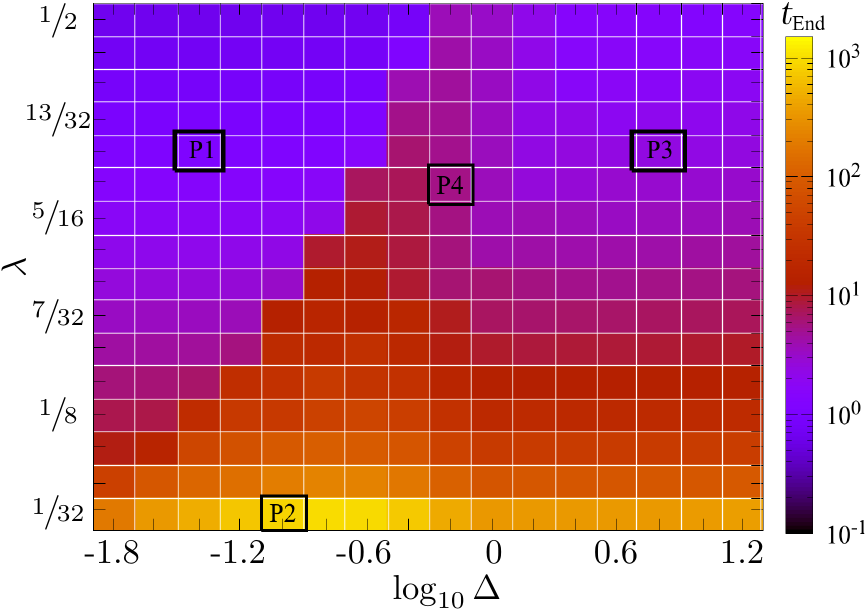}
\caption{
(Color online) Heatmap of the process ending time $t_{\rm End}$ as a function 
of the qubit's energy splitting $\Delta$ and the strength of coupling with its
environment $\lambda$.}
\label{fig:deco-time1}
\end{figure}
%}}}

In Fig.~\ref{fig:deco-time1} we show the process ending time as a function of 
the parameters $\Delta$ and $\lambda$, color coded over the parameter space. 
While we use a linear scale for $\lambda$, $\Delta$ is varied on a log-scale. 
The resulting ending time varies over several
orders of magnitude, so we also use a log-scale for the color mapping. 
While $\lambda$ increases, the $t_{\rm End}$ becomes smaller exponentially 
fast, since for large $\lambda$, the Fermi golden rule approximation 
applies~\cite{CGS14}. It is however 
quite remarkable that for small $\lambda$, the largest $t_{\rm End}$ can be 
found near $\Delta = 0.16$, which is approximately equal to $t_{\rm H}^{-1}$, 
where $t_{\rm H}= 2\pi$ is the Heisenberg time in the random matrix 
environment. In other words, at $\Delta = 0.16$ the period of the Rabi 
oscillation is equal to the Heisenberg time. 

For all non Markovianity measures we choose $t=t_{\rm End}$ unless otherwise 
stated, and we shall thus drop the time dependence. 
% }}}

\subsection{\label{SM} Three measures for non-Markovianity} % {{{
\begin{figure}%{{{
\begin{center}
\includegraphics[width=\columnwidth]{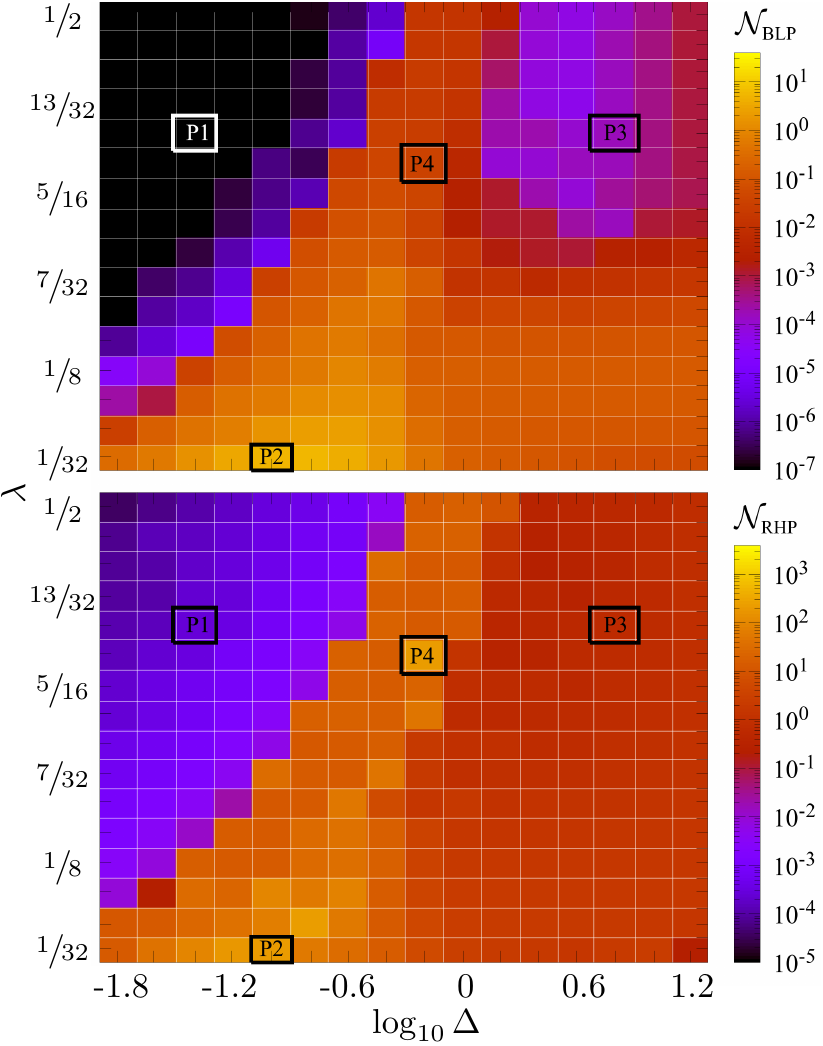}
\end{center}
\caption{(Color online) BLP-measure (upper panel) and RHP-measure 
(lower panel) for non-Markovian dynamics as a function of $\Delta$ and 
$\lambda$. The parameter
region considered is the same as in Fig.~\ref{fig:deco-time1}; again we use a
log-scale in the heatmap representing the values of the NM-measures. Both 
measures, defined in Eqs.~(\ref{eq:IBLPdef}) and~(\ref{eq:IRPLdef}), 
respectively, are taken at $t= t_{\rm End}$. } 
\label{fig:sim5}
\label{fig:sim6}
\end{figure}
%}}}

Analyzing Fig.~\ref{fig:sim5}, we find that both measures reach smallest values 
in the upper left corner of the parameter space, shown. Indeed, due to the 
following argument, we expect the quantum process to be at least close to 
Markovian in this region. The standard prescription for 
deriving a quantum master equation via the Born-Markov approximation consists
in the following steps~\cite{BrePet02}: (i) Couple the central system weakly 
to each of the many degrees of freedom in the environment (Born approximation),
(ii) let the number of degrees of freedom in the environment go to infinity, 
and (iii) assume the environment correlation functions to decay almost 
instantaneously on the time scale of the reduced dynamics (Markov condition). 
In terms of level density and average local level spacing, condition (i) and 
(ii) lead to a wide range in energy with an exponentially high level density,
which means that the perturbation strength will be large as compared to the 
level spacing. This regime is known as the Fermi-golden-rule 
regime~\cite{merzbacher1998quantum}.
Note that in this parameter region, $\Delta \ll 1$, which results in a slow 
system dynamics so that condition (iii) is fulfilled. 

By contrast, for sufficiently small coupling $\lambda < 7/32$ and not too small 
$\Delta$, the dynamics is clearly NM. It is clear that in this region at least 
some of the conditions mentioned above are not fulfilled. The region of 
strongest NM behavior is around $\Delta \approx t_{\rm H}^{-1}$ and small 
values of $\lambda$.

An interesting area is in the upper right region 
of the NM maps in Fig.~\ref{fig:sim5}. There, the BLP measure of 
\Eref{eq:IBLPdef} tends to very small values, while the RHP measure, 
\Eref{eq:IRPLdef}, remains constant. As explained at the end of 
Sec.~\ref{NC}, the RHP criterion is more restrictive than 
the BLP criterion, as it requires complete positivity and not just positivity 
for the intermediate maps. It is thus possible that a given quantum process is 
BLP-Markovian but not RHP-Markovian. Due to residual statistical fluctuations, 
a definite judgement is difficult. 

\begin{figure}% {{{
\begin{center}
\includegraphics[width=\columnwidth]{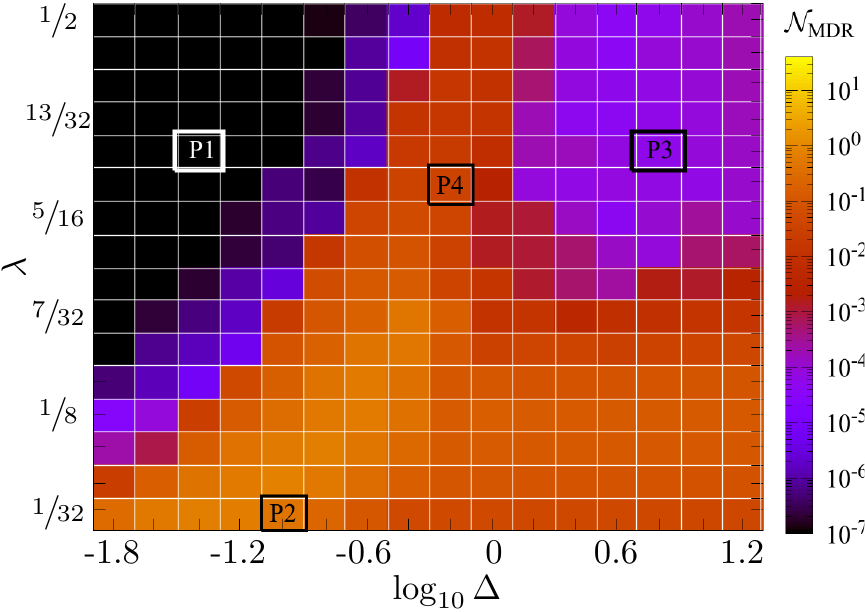}
\end{center}
\caption{(Color online) Maximum distinguishability recovery (MDR). The quantity 
$\mathcal{N}_{\rm MDR}$ as defined in Eq.~(\ref{def:MRtr}), plotted similarly
to the previous two NM-measures in Fig.~\ref{fig:sim5}.}
\label{fig:MRtr}
\end{figure} 
% }}}

Fig.~\ref{fig:MRtr} shows $\mathcal{N}_{\rm MDR}(t)$ as defined in 
Eq.~(\ref{def:MRtr}) up to the ending time $t= t_{\rm End}$ as a function of 
$\Delta$ and $\lambda$. Its behavior is more similar to 
$\mathcal{N}_{\rm BLP}$ than to
$\mathcal{N}_{\rm RHP}$, which may be related to its common origin.
Note though that the boundary between the regions of Markovian and strongly 
non-Markovian behavior is sharper. The region of strong NM is also somewhat 
larger, located in an area parallel to the M-NM boundary, at values for 
$\lambda$ below $1/4$, of about 12 blocks in size. 
Finally, note the region in the upper right corner. There, the MDR measure 
leads to (relatively) lower values than the BLP measure, in distinction to the
RHP measure which reaches much larger values. 
% }}}

\subsection{\label{SS} Robustness and convergence of the NM measures} % {{{
In the previous section, we presented the results of three different measures 
for NM. For our model, we found a wide range of different behavior, depending
on the choice of the parameters, $\Delta$ and $\lambda$. Here, we study the 
robustness and accuracy of the measures in more detail. For that purpose 
we select three points in parameter space, where the behavior of the quantum 
process is quite different: 
\begin{itemize}
\item Point P1 ($\Delta = 10^{-1.4}, \lambda = 3/8$), where the 
  dynamics is Markovian, or at least very close to it.
\item Point P2 ($\Delta = 0.1, \lambda = 1/32$), where it is 
  maximally non-Markovian -- according to the BLP-measure. 
\item Point P3 ($\Delta = 10^{0.8}, \lambda = 3/8$), where the dynamics 
looks like being more NM according to RHP than according to BLP  or
MDR -- compare the two heat maps in Fig.~\ref{fig:sim5}.
\end{itemize}
Finally, we select a additional fourth point for a later remark:
\begin{itemize}
\item Point P4 ($\Delta = 10^{-0.2}, \lambda = 11/32$).
\end{itemize}

\begin{figure} % {{{
\includegraphics[width=\columnwidth]{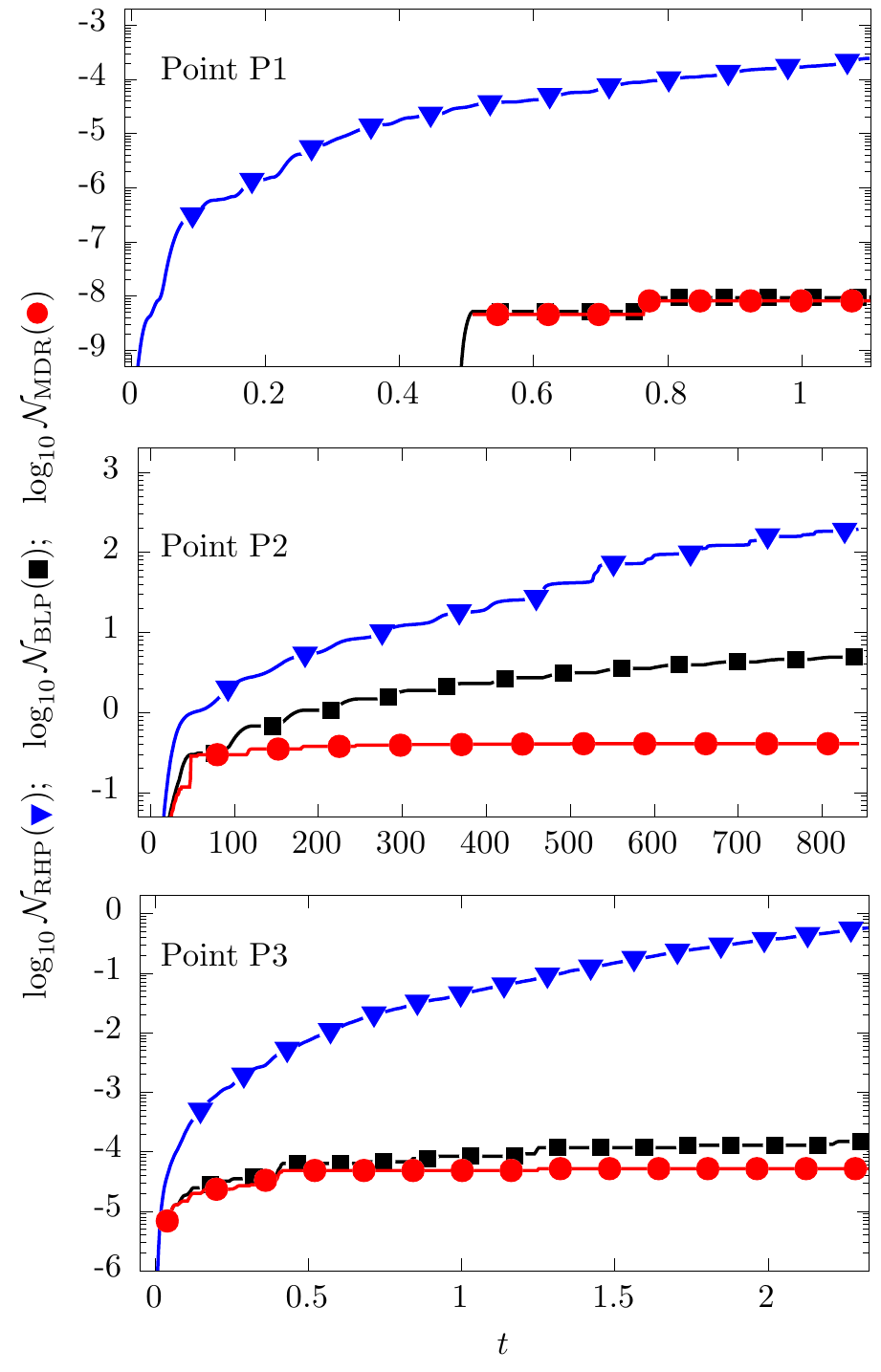}
\caption{(Color online)
NM measures for the three representative points P1, P2, and P3 (for details, 
see the main text), 
from the parametric plane as a function of time.  }
\label{fig:spb-1}
\end{figure}  % }}}

Let us now discuss the numerical results. In Fig.~\ref{fig:spb-1}, the 
NM-measures are shown as a function of time. In other words, we compute the 
NM-measures as if the quantum process would end at time $t$, instead of 
$t_{\rm End}$. It is clear from the definition of all three measures, that they 
must be monotonously increasing: $\mathcal{N}(t_1) \ge \mathcal{N}(t_2)$ 
whenever $t_1 \ge t_2$. 

For P1 (Markovian point; top panel), the RHP-measure (blue line, triangles) 
increases continuously with time -- even though it always remains 
rather small. The other two measures by contrast show only one 
increment at $t\approx 0.5$ and afterwards remain approximately constant
around a value of $10^{-8}$. This makes it difficult to decide unambiguously
whether the process is Markovian or non-Markovian.

For P2 (strongly NM; middle panel), the RHP-measure increases to very large 
values of the order of $10^2$. The BLP measure also increases along the full 
time range up to values of the order $10^0$, while the MDR measure quickly 
saturates at a value of the order of
$10^{-1}$ (note that the MDR measure is by definition limited to values below 
one). The different behavior between BLP and MDR will be discussed below, where
we consider the criteria for contractivity. In any case, all three measures 
clearly show the NM of the process.

For P3 the RHP-measure increases continuously as in the previous cases, 
reaching values of the order of $10^0$. By contrast, the other two measures, 
BLP and MDR remain below a value of the order of $10^{-4}$. This may hint 
towards the possibility that here, the quantum process is P-divisible but not 
CP-divisible.  

\begin{figure} % {{{
\begin{center}
\includegraphics[width=\columnwidth]{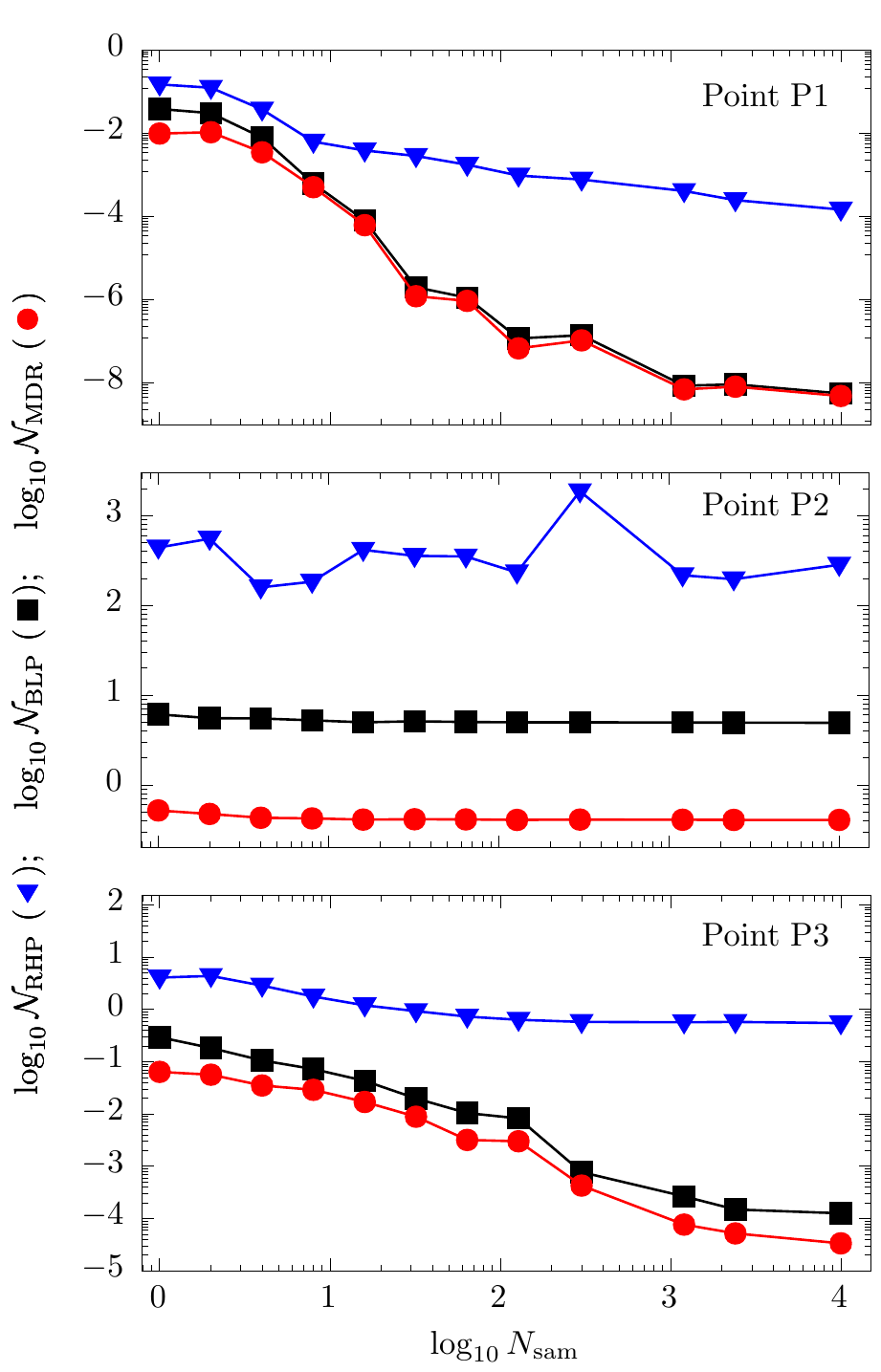}
\end{center}
\caption{(Color online)
NM measures for the same three representative points P1, P2, and P3, as in 
Fig.~\ref{fig:spb-1}, as a function of the sample size. }
\label{fig:brr-1}
\end{figure} 
% }}}

In Fig.~\ref{fig:brr-1}, we plot the NM-measures versus 
the sample size, $N_{\rm sam}$, where we expect that the 
NM-measures approach a limit value for $N_{\rm sam}\to\infty$, the true 
ensemble average. For point P1 (top panel), as the ensemble size increases, 
all measures tend algebraically to zero. For point P2  (middle panel) all 
measures converge to a finite values, whereas for point P3, the measures based 
on distinguishability seem to drop to zero while the one based on divisibility 
attains a finite value. In other words, this result suggests that the dynamics 
is P-divisible but not CP-divisible at that point~\cite{Chr11}.

\subsubsection*{Environment size}
We also experimented with different environment sizes, but the results did not
change significantly. In fact, it can be shown that this may have an effect
at short times only; in general, one would expect finite size effects of
$N_\rme$ at times of the order of $1/N_\rme$ in units of the Heisenberg time. 
These are not our  concern, since we define our model in the limit 
$N_\rme \to\infty$.

\subsubsection*{Non-equivalence of NM measures}
A second interesting question is that of ``quantitative equivalence''. Two
measures $M_1, M_2$ for a physical property may be called (quantitatively) 
equivalent, if and only if 
\[ M_1(A) < M_1(B) \Leftrightarrow M_2(A) < M_2(B) \; , \]
for any two states $A,B$ of some system. For example, if one thermometer 
(calibrated according to the empirical temperature $M_1$) finds that a body $A$ 
is colder than a body $B$, any other thermometer (calibrated according to some 
different empirical temperature $M_2$) should find the same relation. 
Analyzing carefully our results in Fig.~\ref{fig:sim5}, we can 
indeed find pairs of points in parameter space, where the two NM-measures 
$\mathcal{N}_{\rm RHP}$ and $\mathcal{N}_{\rm BLP}$ violate this condition. For 
instance, $\mathcal{N}_{\rm RHP}$ is clearly larger at P4 than at P2, while in 
the case of $\mathcal{N}_{\rm BLP}$ it is just the other way round.
% }}}

\subsection{\label{SC} Time evolution of Markovianity criteria} % {{{
Here, we study the behavior of the two time-local criteria for non-Markovianity 
which are based on divisibility and contractivity, described in Sec.~\ref{N}.

\begin{figure}% {{{ Figura de criterio para la divisibilidad
\begin{center}
\hspace*{-0.15cm}\includegraphics[scale=0.75]{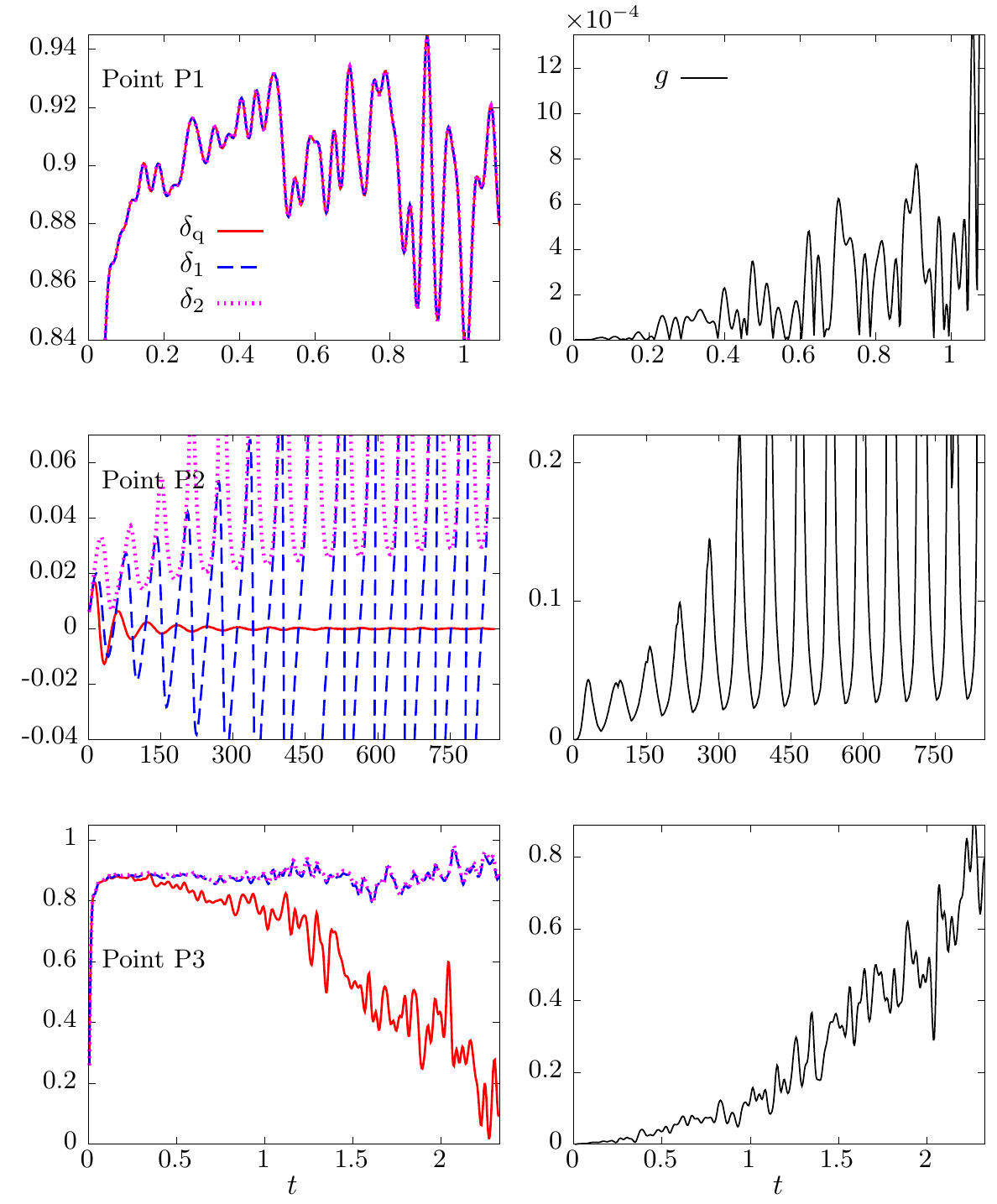}
\end{center}
\caption{(Color online)
Local divisibility criteria as a function of time for the well known three
points from the parametric plane. The left column shows the quantities, 
$\delta_{1,2,q}$ as defined in Eq.~(\ref{def:deltas}), the right column shows 
its corresponding function $g(t)$ as defined in Eq.~(\ref{eq:goft}). For the  
intermediate process to be divisible, the quantities on the left must fulfill
$\delta_2\le \delta_q\le \delta_1$, and the quantity on the right: $g(t) = 0$.
We present this figures for points P1 (top panel), P2 (middle panel), and
P3 (bottom panel). }
\label{fig:sim1}\end{figure} 
% }}}

\begin{figure} % {{{  Figura de criterio para la distinguilibidad, como la 
\begin{center}
\hspace*{-0.15cm}\includegraphics[scale=0.70]{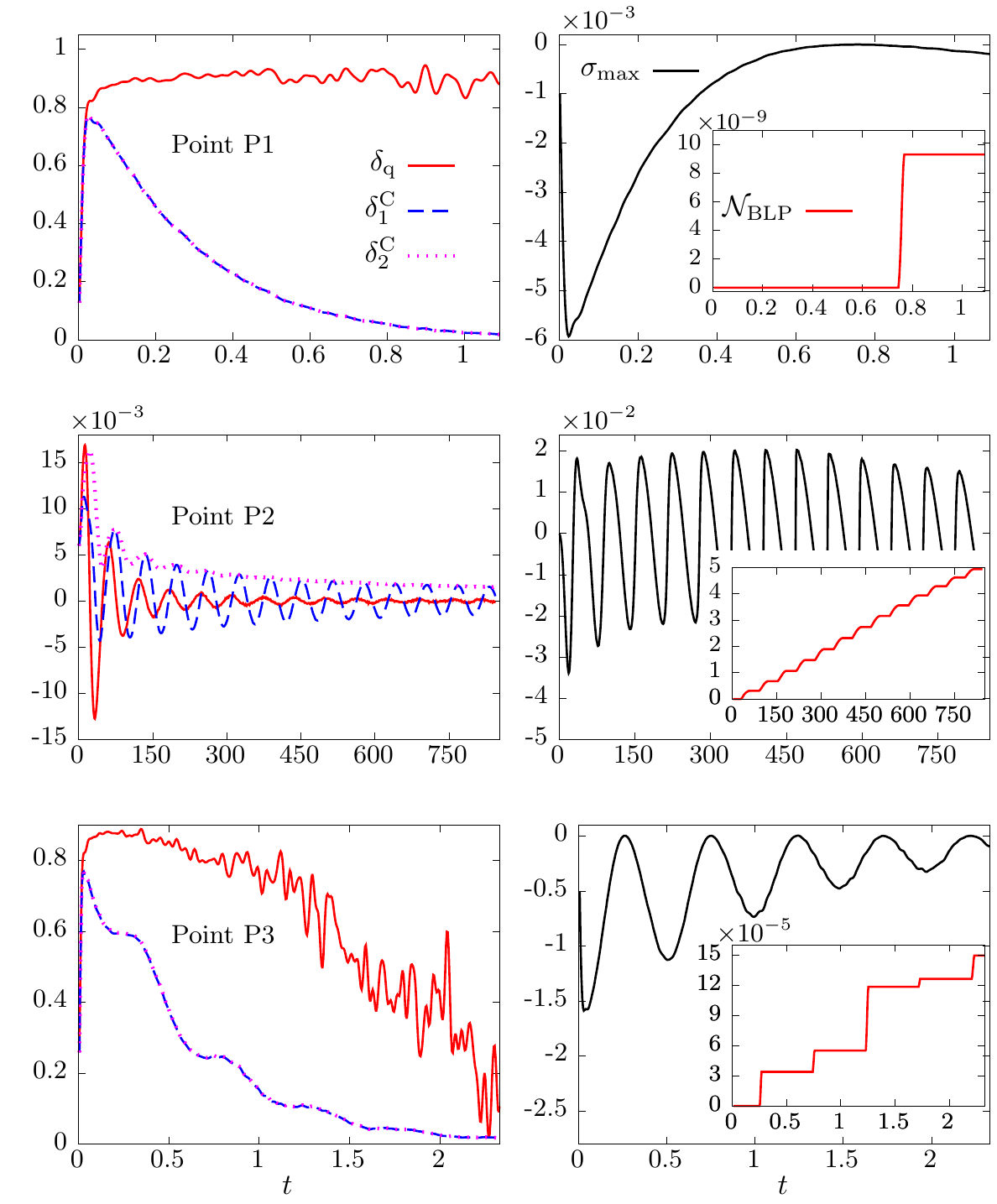}
\end{center}
\caption{(Color online) Local contractivity criteria as a 
function of time. In the left figures, we plot the functions which characterize
the channel, to study how the conditions~\eref{condidistc}, aka $\delta_q \ge
0$ and $\delta_1^{\rm C} \ge \delta_2^{\rm C}$, are fulfilled or not. On the 
right we plot  $\sigma_{\rm max}$, see \eref{eq:sigmamax}, to study how the
measure is build with time, together with its integral (insets).  Again, we 
study points P1, P2 and P3 in the top, middle and bottom panels. }
\label{fig:sim3}\end{figure} % }}}

% Divisibility {{{
For the divisibility, we consider the requirements given
in~\Eref{divcondition} on the one hand, and the condition $g(t)= 0$ 
with $g(t)$ given in \Eref{N:gdefgeneral}.
While formally, both criteria are equivalent [see
\eref{eq:goft}], we will see that in the presence of experimental/statistical
uncertainties, one might be easier to verify than the other. In \Fref{fig:sim1}
we present the aforementioned expressions for points P1, P2 and P3 for times
ranging from zero to $t_{\rm End}$. 

For point P1 (top two panels), we see that the inequalities are saturated, in
the sense that $\delta_{1,2,q}$ are apparently all equal. 
Despite that, the function $g(t)$ is not identically equal to zero, however it 
is small. The inequalities have allowed to correctly identify the point as 
Markovian, in agreement with the analysis of \Fref{fig:brr-1}. In fact, for 
larger ensemble sizes the value of $g(t)$ diminishes. 
 
The behavior of $\delta_{1,2,q}$ and $g(t)$ for point P2 can be seen in the
middle panels of \Fref{fig:sim1}. Here, the different curves corresponding to
$\delta_{1,2,q}$ cross each other in a systematic and ordered fashion,
indicating non-Markovianity beyond statistical fluctuations. This is also seen
in the behaviour of the corresponding $g(t)$, which oscillates regularly around
values of the order of $0.1$. Notice that we identified this point as
displaying non-Markovian behavior, again with the aid of \Fref{fig:brr-1}. 
 
Finally, point P3 is studied in the lower panels of \Fref{fig:sim1}. We can see
that the curve corresponding to $\delta_q$ is not between the ones
corresponding to $\delta_{1,2}$. The conclusions are confirmed by
the behavior of $g(t)$, where one finds notorious fluctuations on top of a 
smooth curve which increases 
systematically as a function of time. Thus, according to the divisibility
criterion, the system is non-Markovian, as concluded from the lower panel 
of \Fref{fig:brr-1}. 
% }}}

% Discucion de contractividad {{{
In the case of the Markovianity measure based on the contractivity of the
process, the condition $\sigma \le 0 $  translates to~\Eref{condidistc} for the 
channels here considered, see \Eref{G:ChoiLam}.
For a given time $t$, a certain initial pair of states 
$\rho_{1,2}^{(\max)}$ will yield the maximum $\mathcal{N}_{\rm BLP}(t)$ as 
defined in \Eref{eq:IBLPdef}.
For these states, one can calculate 
\begin{equation}
\sigma_{\max}= \sigma(\rho_{1}^{(\max)}, \rho_2^{(\max)}, t),
\label{eq:sigmamax}
\end{equation}
and see how the final value of the measure is built with time.  
Notice that $\sigma_{\max}$ is different from the 
derivative of $\mcN_{\rm BLP}(t)$, recall \eref{eq:IBLPdef}, as for each ending 
time $t$, the states that maximize the quantifier are different, whereas 
in \eref{eq:sigmamax} we fix the ending time; however, for $t$ equal to 
the ending time, they coincide. 

On the top left panel we can see that \Eref{condidistc} 
is apparently fulfilled during the whole process, however, since
the two curves for $\delta_{1}^{\rm C}$ and $\delta_{2}^{\rm C}$ lie on top of
each other, the inequality $\delta_{1}^{\rm C} \ge \delta_{2}^{\rm C}$ might be
violated on a smaller scale.
On the top right panel, we see that indeed the measure is close
to zero; for $t< 0.7$ it is numerically zero, and afterwards, close to 
$10^{-8}$. 
From this evidence, we arrive at the conclusion that the point is Markovian, 
with respect to contractivity, in agreement with the same
case studied in Fig.~\ref{fig:sim1}.

The point P2 is analyzed in the middle panels. The function $\delta_{q}$ 
oscillates around zero with decreasing amplitude, while $\delta_2^{\rm C}$ 
provides an upper bound for $\delta_{1}^{\rm C}$. Indeed, whenever this bound
is saturated, $\delta_{q}$ has an node, and $\sigma_{\max}$ has a minimum, 
making the system ``very Markovian''. On the other hand, when the difference 
between $\delta_{1}^{\rm C}$ and $\delta_{2}^{\rm C}$ is largest,  
$\sigma_{\max}$ has a maximum, and the system becomes very non-Markovian. Thus, 
as can be seen on the right, $\sigma_{\max}$ oscillates regularly and with a 
relatively high amplitude around zero. The NM measure, 
$\mathcal{N}_{\rm BLP}$ adds up the areas below the positive parts of 
$\sigma(t)$, which we expect to have a similar behavior as 
$\sigma_{\rm max}(t)$, shown here.  
Therefore, the point P2 shows a genuine non-contractive behavior
and the dynamics is NM in this case.

The point P3 is analyzed in the bottom panels. All three functions $\delta_{q}$
and $\delta_{1,2}^{\rm C}$ display a similar behaviour as for point P1, but for
$\delta_{q}$ we observe stronger statistical fluctuations. In this case,
however, $\sigma_{\max}$ oscillates, and has a maximum in 0.  The measure picks
up small statistical fluctuations which diminish as we increase the sample size 
and may therefore be regarded as spurious. We can conclude that 
the process for point P3 is contractive, in agreement with previous 
conclusions.
% }}}

The criteria provided in Eqs.~(\ref{divcondition}) and~\eqref{condidistc}, 
indeed provide a usefull tool to understand if a certain non-zero 
value for one of the NM measures should be regarded as statistically 
significant or not. In some cases it is helpful
to analyze the behavior of the measures under variation of the sample size in 
order to arrive at the correct decision. 

\section{\label{C} Conclusions} % {{{

In this article we studied a qubit coupled to a generic environment 
modeled by random  matrices. The model Hamiltonian contains a factorizable 
interaction between qubit and environment, and provides the qubit with an
internal dynamics perpendicular (in the Bloch representation) to the one 
induced by the interaction. This induces a channel structure for which we were 
able to derive analytical conditions for several criteria of Markovianity. 
In spite of its simplicity, the model displays rich dynamics in the qubit, 
beyond pure depolarization or dephasing. 

We then applied the criteria to determine for which parameters the model yield
Markovian dynamics in the qubit. We found several difficulties with verifying
Markovianity criteria for the numerical data, which we expect to appear also
in real experimental situations. Fluctuations due to noise, and/or due
to finite sample sizes may contribute notably to the finite value of a
non-Markovianity measure and thereby suggest non-Markovian behavior, whereas
the clean system really is Markovian. An analysis like the present one, where
the ensemble size is increased such that residual fluctuations diminish,
eventually reveals the true behavior.

% \fxnote{Las hace carlos}
% * Ontroducir un modelo de RMT generico y rico, en el sentido en qeu es mas que solo desfazamietno o depolarizacion.
% * Esto se logro haciendo que la interaccion fuera separable, pero no permute con el hamiltoniano del qubit. 
% * Caracterizamos el canal y vimos que tiene dephasing y depolarizacon de manera como indepdntiende. 
% 
% * Estudiammos analiticamente dos criterios de no markovianidad y dimos condiciones sobre el canal para que sean 
% o no markovianos. Obtuvimos las condiciones en terminos de desigualdades de funciones que componen el canal. 
% 
% * Hicimos una exploración numerica del sistema, con una dimension finita y una estadistica finita, estudiando el 
% efecto sobre los criterios. 
% 
% * Definimos un tiempo de decoherencia, y encntramos los regimenes de ????
% 
% para las medidas encontramos reciones donde las dos son muy pequeñoas, 

% }}}
{\em Acknowledgements--} Support by projects % {{{
CONACyT 285754, UNAM-PAPIIT IG100518, Benemérita Universidad
Autónoma de Puebla (BUAP) PRODEP 511-6/2019-4354 and CONACyT 220624-CB 
are acknowledged

% }}}
\begin{appendix}
\begin{widetext}
\section{\label{app:DS} Born series}

In this section, we prove that the quantum channel, describing the dynamics of
the qubit under coupling to the RMT environment, has the form of a 
X-state~\cite{RoMuGr14} as postulated in Eq.~(\ref{G:ChoiLam}). For doing so, 
we use the entire expansion of the evolution of system and environment in a 
Born series. We consider the more general case of an
arbitrary mixed initial state $\varrho_\rme$ in the environment, and only at
the end specialize to the case $\varrho_\rme = \One/N_\rme$.

In the interaction picture, a solution to the Hamiltonian
\[ H_\lambda = \frac{\Delta}{2}\; \sigma_z \otimes\One + \One\otimes H_\rme
      + \lambda\; \sigma_x\otimes V_\rme \; , \]
may be written as
\[ \Psi(t)= U_0(t)\, \chi(t) \quad : \quad \rmi\hbar\, \partial_t\, \chi(t) = 
      \lambda\; U_0(t)^\dagger\, \sigma_x \otimes V_\rme\, U_0(t)\; \chi(t) 
 \; , \quad U_0(t)= \rme^{-\rmi\, \sigma_z t} \otimes 
         \rme^{-\rmi H_\rme t} \; .
 \]
Here, we stick to the convention chosen in the main part of this paper, where
the time variable $t$ measures time in units of the Heisenberg time, which 
results in $\hbar =1$. The echo-operator $M(t)$ describes the evolution of the 
state $\chi(t)$, such that $\chi(t)= M(t)\, \chi(0)$. In the original 
Schr\" odinger picture, it thus holds: 
%\fxnote{no entiendo que sigfnifica el $\Leftrightarrow$ aca}
\[ \Psi(t)= U_0(t)\, M(t)\, \Psi(0) \; , \qquad
   M(t)= U_0(t)^\dagger\; U(t) \; . \]
As a formal solution of the evolution equation in the interaction picture, the
echo operator fulfills the following integral equation
\begin{align}
M(t) &= M(0) - \rmi\lambda\int_0^t\rmd\tau\; 
      \tilde\sigma_x(\tau)\otimes\tilde V_\rme(\tau)\; M(\tau) \notag\\
 &= \sum_{k=0}^\infty \big (-\rmi\lambda\big )^k
     \int\ldots\int_{t > \tau_k > \ldots > \tau_1 > 0}
     \rmd\tau_k\ldots\rmd\tau_1\; 
 \begin{pmatrix} 0 & A(\tau_k) \\ A(\tau_k)^\dagger & 0 \end{pmatrix} \; \ldots
 \begin{pmatrix} 0 & A(\tau_1) \\ A(\tau_1)^\dagger & 0 \end{pmatrix} \; . 
\label{aB:defM}\end{align}
Here, the second line represents the afore mentioned Born series, where the 
interaction has been written in block-matrix notation:
\[ \tilde\sigma_x(\tau)\otimes\tilde V_\rme(\tau) =
   \begin{pmatrix} 0 & \rme^{\rmi\Delta\tau}\, \tilde V_\rme(\tau)\\
      \rme^{-\rmi\Delta\tau}\, \tilde V_\rme & 0 \end{pmatrix}
 = \begin{pmatrix} 0 & A(\tau) \\ A(\tau)^\dagger & 0 \end{pmatrix} \; . \]
Here, we will calculate the average of
$\tilde\varrho(t) = M(t)\, \varrho_{\rm c}\otimes\varrho_\rme\, M(t)^\dagger$,
with respect to the random matrix ensemble for $V_\rme$. This defines the
quantum map $\Lambda_t$ as 
\begin{align}
 \Lambda_t \quad :\quad \varrho_\rmc(0) \to 
   \varrho_\rmc(t) = {\rm tr}_\rme\big [ U(t)\, 
      \varrho_\rmc(0)\otimes\varrho_\rme\, U(t)^\dagger \big ]
 = \begin{pmatrix} \rme^{-\rmi\Delta t/2} & 0 \\ 
      0 & \rme^{\rmi\Delta t/2}\end{pmatrix}\; {\rm tr}_\rme\big [\,
   \tilde\varrho(t)\, \big ]\;
   \begin{pmatrix} \rme^{\rmi\Delta t/2} & 0 \\ 
      0 & \rme^{-\rmi\Delta t/2}\end{pmatrix}\; , 
\label{aB:LambdaSchPic}\end{align}
and thereby its Choi-matrix representation, in Eq.~(\ref{G:ChoiLam}). Below, we 
will also consider this quantum map in the interaction picture, defined as
\begin{align}
 \tilde\Lambda_t \quad :\quad \varrho_\rmc(0) \to 
   \tilde\varrho_\rmc(t) = {\rm tr}_\rme\big [\, \tilde\varrho(t)\, \big ]\; .
\label{aB:LambdaIntPic}\end{align}

\paragraph{Simplified notation}
For our purpose, it is convenient to introduce the following more compact 
notation for the multi-dimensional time integrals:
\begin{align}
\mathcal{I}^0(\bm\tau) &= 1 \; , \qquad
\mathcal{I}^1(\bm\tau) = \mathcal{I}(\tau) = \int_0^t\rmd\tau \; , \qquad
\mathcal{I}^2(\bm\tau) = \mathcal{I}(\tau_2,\tau_1) 
  = \int_{t > \tau_2 > \tau_1 > 0}\rmd^2\bm\tau
  = \int_{t > \tau_2 > \tau_1 > 0}\rmd\tau_2\, \rmd\tau_1\notag\\
\mathcal{I}^k(\bm\tau) &= \int_{t > \tau_k > \ldots > \tau_1 > 0}
     \rmd\tau_k\ldots\rmd\tau_1 \; .
\end{align}
Note that we enumerate the different time variables from smallest to largest 
starting at the time closest to zero. Next, we will introduce an independent
notation for the integrands. Let us consider the caso of an even number of 
product terms, first:
\[ \prod_{j=1}^{2k} 
   \begin{pmatrix} 0 & A(\tau_j) \\ A(\tau_j)^\dagger & 0 \end{pmatrix}
 = \begin{pmatrix} P_{2k}(\bm\tau) & 0 \\ 0 & Q_{2k}(\bm\tau)
   \end{pmatrix} \; , \qquad
 A(\tau)= \rme^{\rmi\Delta\tau}\; \tilde V_\rme(\tau) \; , \qquad
 A(\tau)^\dagger = \rme^{-\rmi\Delta\tau}\; \tilde V_\rme(\tau) 
                 = \rme^{-2\rmi\Delta\tau}\; A(\tau)\; , \]
since $\tilde V_\rme(\tau)$ is Hermitian. Note that the product terms on the 
LHS of the first equation, must be ordered according to decreassing time 
arguments, just as in Eq.~(\ref{aB:defM}). From an explicit computation we find
\begin{align}
P_{2k}(\bm\tau) &= A(\tau_{2k})\; A(\tau_{2k-1})^\dagger\; A(\tau_{2k-2})\;
   A(\tau_{2k-3})^\dagger\; \ldots\; A(\tau_2)\; A(\tau_1)^\dagger
 = \exp\big [ -\rmi\Delta\, {\textstyle\sum_{m=1}^{2k}} (-1)^m\, \tau_m\, 
      \big ]\; \prod_{m=1}^{2k} \tilde V(\tau_m)  \notag\\
Q_{2k}(\bm\tau) &= A(\tau_{2k})^\dagger\; A(\tau_{2k-1})\; 
   A(\tau_{2k-2})^\dagger\; A(\tau_{2k-3})\; \ldots\; A(\tau_2)^\dagger\; 
   A(\tau_1)
 = \exp\big [\, \rmi\Delta\, {\textstyle\sum_{m=1}^{2k}} (-1)^m\, \tau_m\, 
      \big ]\; \prod_{m=1}^{2k} \tilde V(\tau_m)\; .
\end{align}
Note that the product terms must be ordered such that time increases from right
to left. For an odd number of terms:
\begin{align}
\prod_{j=1}^{2k+1} 
   \begin{pmatrix} 0 & A(\tau_j) \\ A(\tau_j)^\dagger & 0 \end{pmatrix}
 &= \begin{pmatrix} 0 & A(\tau_{2k+1}) \\ A(\tau_{2k+1})^\dagger & 0 
   \end{pmatrix}
 \begin{pmatrix} P_{2k}(\bm\tau) & 0 \\ 0 & Q_{2k}(\bm\tau)
   \end{pmatrix} 
 = \begin{pmatrix} 0 & A(\tau_{2k+1})\, Q_{2k}(\bm\tau) \\ 
      A(\tau_{2k+1})^\dagger\, P_{2k}(\bm\tau) & 0 \end{pmatrix} \notag\\
 &= \begin{pmatrix} 0 & Q_{2k+1}(\bm\tau) \\ 
      P_{2k+1}(\bm\tau) & 0 \end{pmatrix} \; .
\end{align}
With this, we may write for the echo operator 
\begin{align}
M(t)= \sum_{k=0}^\infty \Big\{ \, (-\rmi\lambda)^{2k}\; 
   \mathcal{I}^{2k}(\bm\tau)\;
   \begin{pmatrix} P_{2k}(\bm\tau) & 0 \\ 0 & Q_{2k}(\bm\tau)\end{pmatrix}
 + (-\rmi\lambda)^{2k+1}\; \mathcal{I}^{2k+1}(\bm\tau)\;
   \begin{pmatrix} 0 & Q_{2k+1}(\bm\tau) \\ P_{2k+1}(\bm\tau) & 0\end{pmatrix}
\, \Big\}\; .
\end{align}

\paragraph{Ensemble averaged quantum channel}
Averaging over the random matrix $V_\rme$ implies that only such terms survive,
which contain an even power of matrices $A(\tau_k)$ and $A(\sigma_{k'})$. This
means that the indices of summation must either be both even or both odd.
Therefore,
\begin{align}
\tilde\varrho(t) &= \sum_{k,k'=0}^\infty (-\rmi\lambda)^{2k}\; 
   \mathcal{I}^{2k}(\bm\tau)\;
   \begin{pmatrix} P_{2k}(\bm\tau) & 0 \\ 0 & Q_{2k}(\bm\tau)\end{pmatrix}\;
   \varrho_\rmc\otimes\varrho_\rme\;
   (\rmi\lambda)^{2k'}\; 
   \mathcal{I}^{2k'}(\bm\tau)\;
   \begin{pmatrix} P_{2k'}(\bm\tau)^\dagger & 0 \\ 0 & Q_{2k'}(\bm\tau)^\dagger
   \end{pmatrix}\notag\\
&\qquad + \sum_{k,k'=0}^\infty
  (-\rmi\lambda)^{2k+1}\; \mathcal{I}^{2k+1}(\bm\tau)\;
   \begin{pmatrix} 0 & Q_{2k+1}(\bm\tau) \\ P_{2k+1}(\bm\tau) & 0\end{pmatrix}\;
   \varrho_\rmc\otimes\varrho_\rme\;
  (\rmi\lambda)^{2k'+1}\; \mathcal{I}^{2k'+1}(\bm\tau)\;
   \begin{pmatrix} 0 & Q_{2k'+1}(\bm\tau)^\dagger \\ 
      P_{2k'+1}(\bm\tau)^\dagger & 0\end{pmatrix}\notag\\
 &= \sum_{k,k'=0}^\infty (-\lambda^2)^{k+k'}\; \Big\{ \,
   \mathcal{I}^{2k}(\bm\tau)\; \mathcal{I}^{2k'}(\bm\sigma)\;
   \begin{pmatrix} P_{2k}(\bm\tau) & 0 \\ 0 & Q_{2k}(\bm\tau)\end{pmatrix}\;
   \varrho_\rmc\otimes\varrho_\rme\;
   \begin{pmatrix} P_{2k'}(\bm\sigma)^\dagger & 0 \\ 
      0 & Q_{2k'}(\bm\sigma)^\dagger \end{pmatrix}\notag\\
&\qquad + \lambda^2\; \mathcal{I}^{2k+1}(\bm\tau)\; 
   \mathcal{I}^{2k'+1}(\bm\sigma)\;
   \begin{pmatrix} 0 & Q_{2k+1}(\bm\tau) \\ P_{2k+1}(\bm\tau) & 0\end{pmatrix}\;
   \varrho_\rmc\otimes\varrho_\rme\;
   \begin{pmatrix} 0 & Q_{2k'+1}(\bm\sigma)^\dagger \\ 
      P_{2k'+1}(\bm\sigma)^\dagger & 0\end{pmatrix}\; \Big\}
\end{align}
For the quantum channel in the interaction picture, Eq.~(\ref{aB:LambdaIntPic})
it is now easily verified that
\begin{align}
\tilde\Lambda_t[\, |0\ra\la 0|\, ] &= 
  \sum_{k,k'=0}^\infty (-\lambda^2)^{k+k'}\; {\rm tr}_\rme\Big\{ \,
   \mathcal{I}^{2k}(\bm\tau)\; \mathcal{I}^{2k'}(\bm\sigma)\;
   \begin{pmatrix} P_{2k}(\bm\tau) & 0 \\ 0 & Q_{2k}(\bm\tau)\end{pmatrix}\;
   \begin{pmatrix} \varrho_\rme & 0 \\ 0 & 0\end{pmatrix}\;
   \begin{pmatrix} P_{2k'}(\bm\sigma)^\dagger & 0 \\ 
      0 & Q_{2k'}(\bm\sigma)^\dagger \end{pmatrix}\notag\\
&\qquad + \lambda^2\; \mathcal{I}^{2k+1}(\bm\tau)\; 
   \mathcal{I}^{2k'+1}(\bm\sigma)\;
   \begin{pmatrix} 0 & Q_{2k+1}(\bm\tau) \\ P_{2k+1}(\bm\tau) & 0\end{pmatrix}\;
   \begin{pmatrix} \varrho_\rme & 0 \\ 0 & 0\end{pmatrix}\;
   \begin{pmatrix} 0 & Q_{2k'+1}(\bm\sigma)^\dagger \\ 
      P_{2k'+1}(\bm\sigma)^\dagger & 0\end{pmatrix}\; \Big\} \notag\\
 &= \sum_{k,k'=0}^\infty (-\lambda^2)^{k+k'}\; \Big\{ \,
   \mathcal{I}^{2k}(\bm\tau)\; \mathcal{I}^{2k'}(\bm\sigma)\;
   \begin{pmatrix} 
     P_{2k}(\bm\tau)\, \varrho_\rme\, P_{2k'}(\bm\sigma)^\dagger & 0 \\ 
     0 & 0\end{pmatrix}\notag\\
&\qquad + \lambda^2\; \mathcal{I}^{2k+1}(\bm\tau)\; 
   \mathcal{I}^{2k'+1}(\bm\sigma)\;
   \begin{pmatrix} 0 & 0 \\ 
     0 & P_{2k+1}(\bm\tau)\, \varrho_\rme\, Q_{2k'+1}(\bm\sigma)^\dagger
   \end{pmatrix}\; \Big\} \; . 
\end{align}
This and Eq.~(\ref{aB:LambdaSchPic}) then show that 
$\Lambda_t[\, |0\ra\la 0|\, ]$ is indeed of the form postulated in 
Eq.~(\ref{G:ChoiLam}), and it yields the following expressions for $r(t)$:
\begin{align}
 r(t) &= \sum_{k,k'=0}^\infty (-\lambda^2)^{k+k'}\; 
   \mathcal{I}^{2k}(\bm\tau)\; \mathcal{I}^{2k'}(\bm\sigma)\;
   {\rm tr}\big [\, P_{2k}(\bm\tau)\, \varrho_\rme\, P_{2k'}(\bm\sigma)^\dagger
      \, \big ] \\
1 - r(t) &= -\sum_{k,k'=0}^\infty (-\lambda^2)^{k+k'+1}\; 
   \mathcal{I}^{2k+1}(\bm\tau)\; \mathcal{I}^{2k'+1}(\bm\sigma)\;
   {\rm tr}\big [\, P_{2k+1}(\bm\tau)\, \varrho_\rme\, 
      Q_{2k'+1}(\bm\sigma)^\dagger\, \big ] \; .
\end{align}
The second equation results from the fact that the reduced evolution of the 
qubit conserves the trace. Naturally, it is difficult to prove this directly,
from the expressions derived here. Let us now consider
$\Lambda_t[\, |1\ra\la 1|\, ]$. In this case, as in the previous one, 
$\Lambda_t[\, |1\ra\la 1|\, ] = \tilde\Lambda_t[\, |1\ra\la 1|\, ]$, and we
find
\begin{align}
\Lambda_t[\, |1\ra\la 1|\, ] &= 
  \sum_{k,k'=0}^\infty (-\lambda^2)^{k+k'}\; {\rm tr}_\rme\Big\{ \,
   \mathcal{I}^{2k}(\bm\tau)\; \mathcal{I}^{2k'}(\bm\sigma)\;
   \begin{pmatrix} P_{2k}(\bm\tau) & 0 \\ 0 & Q_{2k}(\bm\tau)\end{pmatrix}\;
   \begin{pmatrix} 0 & 0 \\ 0 & \varrho_\rme\end{pmatrix}\;
   \begin{pmatrix} P_{2k'}(\bm\sigma)^\dagger & 0 \\ 
      0 & Q_{2k'}(\bm\sigma)^\dagger \end{pmatrix}\notag\\
&\qquad + \lambda^2\; \mathcal{I}^{2k+1}(\bm\tau)\; 
   \mathcal{I}^{2k'+1}(\bm\sigma)\;
   \begin{pmatrix} 0 & Q_{2k+1}(\bm\tau) \\ P_{2k+1}(\bm\tau) & 0\end{pmatrix}\;
   \begin{pmatrix} 0 & 0 \\ 0 & \varrho_\rme\end{pmatrix}\;
   \begin{pmatrix} 0 & Q_{2k'+1}(\bm\sigma)^\dagger \\ 
      P_{2k'+1}(\bm\sigma)^\dagger & 0\end{pmatrix}\; \Big\} \notag\\
 &= \sum_{k,k'=0}^\infty (-\lambda^2)^{k+k'}\; \Big\{ \,
   \mathcal{I}^{2k}(\bm\tau)\; \mathcal{I}^{2k'}(\bm\sigma)\;
   \begin{pmatrix} 
     0 & 0 \\ 0 & Q_{2k}(\bm\tau)\, \varrho_\rme\, Q_{2k'}(\bm\sigma)^\dagger
   \end{pmatrix}\notag\\
&\qquad + \lambda^2\; \mathcal{I}^{2k+1}(\bm\tau)\; 
   \mathcal{I}^{2k'+1}(\bm\sigma)\;
   \begin{pmatrix} 
      Q_{2k+1}(\bm\tau)\, \varrho_\rme\, P_{2k'+1}(\bm\sigma)^\dagger & 0\\
      0 & 0 \end{pmatrix}\; \Big\} 
 = \begin{pmatrix} 1 - \tilde r(t) & 0\\ 0 & \tilde r(t)\end{pmatrix} \; .
\end{align}
Again, the resulting qubit state is diagonal, however unless we specialize to
the case $\varrho_\rme = \One/N$, the function $\tilde r(t)$ is different from
$r(t)$ corresponding to the previous case.
\begin{align}
\tilde r(t) &= \sum_{k,k'=0}^\infty (-\lambda^2)^{k+k'}\;
   \mathcal{I}^{2k}(\bm\tau)\; \mathcal{I}^{2k'}(\bm\sigma)\;
   {\rm tr}\big [\, 
      Q_{2k}(\bm\tau)\, \varrho_\rme\, Q_{2k'}(\bm\sigma)^\dagger\, \big ]\; ,
\notag\\
1- \tilde r(t) &= -\sum_{k,k'=0}^\infty (-\lambda^2)^{k+k'+1}\; 
 \mathcal{I}^{2k+1}(\bm\tau)\; \mathcal{I}^{2k'+1}(\bm\sigma)\;
   {\rm tr}\big [\, 
   Q_{2k+1}(\bm\tau)\, \varrho_\rme\, P_{2k'+1}(\bm\sigma)^\dagger\, \big ]\; .
\end{align}
We continue with the off-diagonal blocks of the Choi-matrix representation of
the quantum channel:
\begin{align}
\tilde \Lambda_t[\, |1\ra\la 0|\, ] &= 
  \sum_{k,k'=0}^\infty (-\lambda^2)^{k+k'}\; {\rm tr}_\rme\Big\{ \,
   \mathcal{I}^{2k}(\bm\tau)\; \mathcal{I}^{2k'}(\bm\sigma)\;
   \begin{pmatrix} P_{2k}(\bm\tau) & 0 \\ 0 & Q_{2k}(\bm\tau)\end{pmatrix}\;
   \begin{pmatrix} 0 & 0 \\ \varrho_\rme & 0\end{pmatrix}\;
   \begin{pmatrix} P_{2k'}(\bm\sigma)^\dagger & 0 \\ 
      0 & Q_{2k'}(\bm\sigma)^\dagger \end{pmatrix}\notag\\
&\qquad + \lambda^2\; \mathcal{I}^{2k+1}(\bm\tau)\; 
   \mathcal{I}^{2k'+1}(\bm\sigma)\;
   \begin{pmatrix} 0 & Q_{2k+1}(\bm\tau) \\ P_{2k+1}(\bm\tau) & 0\end{pmatrix}\;
   \begin{pmatrix} 0 & 0 \\ \varrho_\rme & 0\end{pmatrix}\;
   \begin{pmatrix} 0 & Q_{2k'+1}(\bm\sigma)^\dagger \\ 
      P_{2k'+1}(\bm\sigma)^\dagger & 0\end{pmatrix}\; \Big\} \notag\\
 &= \sum_{k,k'=0}^\infty (-\lambda^2)^{k+k'}\; \Big\{ \,
   \mathcal{I}^{2k}(\bm\tau)\; \mathcal{I}^{2k'}(\bm\sigma)\;
   \begin{pmatrix} 
     0 & 0 \\ Q_{2k}(\bm\tau)\, \varrho_\rme\, P_{2k'}(\bm\sigma)^\dagger & 0
   \end{pmatrix}\notag\\
&\qquad + \lambda^2\; \mathcal{I}^{2k+1}(\bm\tau)\; 
   \mathcal{I}^{2k'+1}(\bm\sigma)\;
   \begin{pmatrix} 
      0 & Q_{2k+1}(\bm\tau)\, \varrho_\rme\, Q_{2k'+1}(\bm\sigma)^\dagger\\
      0 & 0 \end{pmatrix}\; \Big\} \; . 
\end{align}
This result confirms again the general X-state structure of the 
Choi-representation of our quantum channel. In terms of the parametrization in 
Eq.~(\ref{G:ChoiLam}), we find:
\begin{align}
z_1(t) &= \rme^{\rmi\Delta t}\; \sum_{k,k'=0}^\infty (-\lambda^2)^{k+k'}\;
   \mathcal{I}^{2k}(\bm\tau)\; \mathcal{I}^{2k'}(\bm\sigma)\;
   {\rm tr}\big [\, Q_{2k}(\bm\tau)\, \varrho_\rme\, P_{2k'}(\bm\sigma)^\dagger
      \, \big ] \notag\\
z_2(t)^* &= \rme^{-\rmi\Delta t}\;
   \sum_{k,k'=0}^\infty (-\lambda^2)^{k+k'}\; \lambda^2\;
   \mathcal{I}^{2k+1}(\bm\tau)\; \mathcal{I}^{2k'+1}(\bm\sigma)\;
   {\rm tr}\big [\, Q_{2k+1}(\bm\tau)\, \varrho_\rme\, 
      Q_{2k'+1}(\bm\sigma)^\dagger \, \big ] \; ,
\label{aB:z1z2Res}\end{align}
where the phases $\rme^{\pm\rmi\Delta t}$ arise from returing to the 
Schr\" odinger picture, according to Eq.~(\ref{aB:LambdaSchPic}). Finally,
\begin{align}
\tilde \Lambda_t[\, |0\ra\la 1|\, ] &= 
  \sum_{k,k'=0}^\infty (-\lambda^2)^{k+k'}\; {\rm tr}_\rme\Big\{ \,
   \mathcal{I}^{2k}(\bm\tau)\; \mathcal{I}^{2k'}(\bm\sigma)\;
   \begin{pmatrix} P_{2k}(\bm\tau) & 0 \\ 0 & Q_{2k}(\bm\tau)\end{pmatrix}\;
   \begin{pmatrix} 0 & \varrho_\rme\\ 0 & 0\end{pmatrix}\;
   \begin{pmatrix} P_{2k'}(\bm\sigma)^\dagger & 0 \\ 
      0 & Q_{2k'}(\bm\sigma)^\dagger \end{pmatrix}\notag\\
&\qquad + \lambda^2\; \mathcal{I}^{2k+1}(\bm\tau)\; 
   \mathcal{I}^{2k'+1}(\bm\sigma)\;
   \begin{pmatrix} 0 & Q_{2k+1}(\bm\tau) \\ P_{2k+1}(\bm\tau) & 0\end{pmatrix}\;
   \begin{pmatrix} 0 & \varrho_\rme \\ 0 & 0\end{pmatrix}\;
   \begin{pmatrix} 0 & Q_{2k'+1}(\bm\sigma)^\dagger \\ 
      P_{2k'+1}(\bm\sigma)^\dagger & 0\end{pmatrix}\; \Big\} \notag\\
 &= \sum_{k,k'=0}^\infty (-\lambda^2)^{k+k'}\; \Big\{ \,
   \mathcal{I}^{2k}(\bm\tau)\; \mathcal{I}^{2k'}(\bm\sigma)\;
   \begin{pmatrix} 
     0 & P_{2k}(\bm\tau)\, \varrho_\rme\, Q_{2k'}(\bm\sigma)^\dagger\\
     0 & 0 \end{pmatrix}\notag\\
&\qquad + \lambda^2\; \mathcal{I}^{2k+1}(\bm\tau)\; 
   \mathcal{I}^{2k'+1}(\bm\sigma)\;
   \begin{pmatrix} 
      0 & 0 \\
      P_{2k+1}(\bm\tau)\, \varrho_\rme\, P_{2k'+1}(\bm\sigma)^\dagger & 0
   \end{pmatrix}\; \Big\} \; .
\end{align}
For any Hermiticity conserving linear map, the Choi-representation itself must 
be Hermitian, in particular also for our quantum channel, as can be seen from
its definition in terms of the reduced dynamics in Eq.~(\ref{G:reducedDyn}).
This implies that %  In our case, the definition of the quantum channel reduced dynamics 
\begin{align}
z_1(t)^* &= \rme^{-\rmi\Delta t}\; \sum_{k,k'=0}^\infty (-\lambda^2)^{k+k'}\;
   \mathcal{I}^{2k}(\bm\tau)\; \mathcal{I}^{2k'}(\bm\sigma)\;
   {\rm tr}\big [\, P_{2k}(\bm\tau)\, \varrho_\rme\, Q_{2k'}(\bm\sigma)^\dagger
      \, \big ] \notag\\
z_2(t) &= \rme^{\rmi\Delta t}\;
   \sum_{k,k'=0}^\infty (-\lambda^2)^{k+k'}\; \lambda^2\;
   \mathcal{I}^{2k+1}(\bm\tau)\; \mathcal{I}^{2k'+1}(\bm\sigma)\;
   {\rm tr}\big [\, P_{2k+1}(\bm\tau)\, \varrho_\rme\, 
      P_{2k'+1}(\bm\sigma)^\dagger \, \big ] \; .
\label{aB:z1z2Res2}\end{align}
In the case of $z_1(t)$ the equivalence of the Eqs.~(\ref{aB:z1z2Res}) 
and~(\ref{aB:z1z2Res2}) is rather obvious. One simply has to exchange the 
variable names $k$ and $k'$ along with $\bm\tau$ and $\bm\sigma$. In the case
of $z_2(t)$, we do not see any simple way of proving the equivalence in terms
of the present approach.

\paragraph{Collecting results for the case $\varrho_\rme = \One/N$}
From the considerations in the previous paragraph, we found that the 
Choi-matrix representation of the quantum channel defined by the
Eqs.~(\ref{G:reducedDyn}) and~(\ref{aB:LambdaSchPic}) is given by
\[ C_{\Lambda_t} = \begin{pmatrix}
      r & 0 & 0 & z_1^*\\
      0 & 1-r & z_2 & 0\\
      0 & z_2^* & 1-\tilde r & 0\\
      z_1 & 0 & 0 & \tilde r\end{pmatrix}\; . \]
The general X-state structure (i.e. all the zeros in this matrix) follows 
directly from the considerations in this section. For simplicity we omitted the
time argument in this representation. The functions $r(t)$, $z_1(t)$, $z_2(t)$,
and $\tilde r(t)$ are as defined above. Some of the dependencies between
these matrix elements could not be proven within our derivation, but are valid
due to elementary properties of the evolution equation~(\ref{G:reducedDyn}).
This is the case for trace conservation (diagonal blocks) and Hermiticity
(off-diagonal blocks). 

As a last point, we show that for $\varrho_\rme = \One/N_\rme$, it holds that
$r(t) = \tilde r(t)$. In this case, and for $\varrho_\rme(0) = \One_2$, we 
find from Eq.~(\ref{G:reducedDyn}):
\begin{align}
\Lambda_t[\One] = \frac{1}{N_\rme}\; {\rm tr}_\rme\big [\, 
   \rme^{-\rmi H_\lambda t}\; \rme^{\rmi H_\lambda t}\, \big ] = \One \; .
\label{aB:unital}\end{align} 
This means that no matte if we perform an ensemble average or not, the 
resulting quantum channel is unital (it maps the identity onto itself). This in
turn implies:
\begin{align}
\Lambda_t[\One] = \Lambda_t[\, |0\ra\la 0|\, ] + \Lambda_t[\, |1\ra\la 1|\, ]
 = \begin{pmatrix} r(t) & 0\\ 0 & 1-r(t)\end{pmatrix} 
 + \begin{pmatrix} 1 - \tilde r(t) & 0\\ 0 & \tilde r(t)\end{pmatrix} 
 = \begin{pmatrix} 1 & 0\\ 0 & 1\end{pmatrix}
\quad\Leftrightarrow\quad r(t) = \tilde r(t)\; .
\end{align}
\end{widetext}

\section{\label{app:res} Two representations of a quantum channel}

%\paragraph{Stuff from the main part -- possible repetition}
We will use two different representations. (i) The ``super operator''
representation which is nothing else than the standard matrix representation of
a linear map on a vector space. This representation is easy to read directly
from the evolution of a standard set of states; in fact, we shall construct it
in that way.  (ii) The Choi-matrix representation in which  verifying inherent
quantum channel properties is easier. \par
% }}}
%% General theory {{{
For the super operator representation, we represent the density matrices as
column vectors, in the so-called ``anti-lexicographical''
ordering~\cite{BenZyc06}.
%That means that we concatenate the columns of the density
%matrix, to construct a column vector. 
For a single qubit, we have
\begin{equation}
\varrho =
\begin{pmatrix} 
\varrho_{00} & \varrho_{01}\\ 
\varrho_{10} & \varrho_{11}
\end{pmatrix}\quad \longleftrightarrow\quad
\vec{\varrho}=
\begin{pmatrix} 
\varrho_{00}\\ 
\varrho_{10}\\ 
\varrho_{01} \\
\varrho_{11}
\end{pmatrix}.
\end{equation}
This fixes the matrix representation $L_t$ of the map $\Lambda_t$, since the
columns of $L_t$ must be the images of the canonical basis vectors. Thus, with
the condensed form
$\Lambda[ij]_{kl} \equiv \la k|\, \Lambda_t[\, |i\ra\la j|\, ]\, |l\ra$, we
obtain
\begin{equation}
L_t = \begin{pmatrix}
   \Lambda[00]_{00} & \Lambda[10]_{00} & \Lambda[01]_{00} & \Lambda[11]_{00}\\
   \Lambda[00]_{10} & \Lambda[10]_{10} & \Lambda[01]_{10} & \Lambda[11]_{10}\\
   \Lambda[00]_{01} & \Lambda[10]_{01} & \Lambda[01]_{01} & \Lambda[11]_{01}\\
   \Lambda[00]_{11} & \Lambda[10]_{11} & \Lambda[01]_{11} & \Lambda[11]_{11}
 \end{pmatrix}.
\label{G:SuperMRep}
\end{equation}
The super operator representation has the convenient property, that the
representation $L$ of the composition of two quantum maps,
$\Lambda = \Lambda_2 \circ \Lambda_1$ (where $\Lambda_2$ is applied to the
result of $\Lambda_1$) is simply given by the matrix product of the
representations $L_1$ and $L_2$ of the individual maps: $L = L_2\, L_1$.

\subsection{\label{aR} Quantum map for intermediate time steps}
%
%Comparing to the Eqs.~(\ref{G:SuperMRep}) and~(\ref{G:ChoiRep}), we obtain
We know that the superopertor has the following form
\begin{equation}
L_t = \begin{pmatrix} r & 0 & 0 & 1-r\\
         0 & z_1 & z_2 & 0\\
         0 & z_2^* & z_1^* & 0\\
         1-r & 0 & 0 & r\end{pmatrix},
\label{eq:lt}
\end{equation}
where $r$ and $z_{1,2}$ are functions on time. 

For given quantum maps $\Lambda_t$ and $\Lambda_{t+\eps}$, we compute the
quantum map which takes states $\varrho(t)$ form time $t$ to $t+\eps$. For the
moment we assume $\eps>0$ to be finite. The central question is whether 
$\Lambda_{t+\eps,t}=\Lambda_{t+\eps}\circ\Lambda_{t}^{-1}$ is a CP-map, or 
not. While the super-operator representation  $L_t$ is appropriate to compute
the composition of $\Lambda_{t+\eps}$ and $\Lambda_{t}^{-1}$, the Choi 
representation is needed for verifying the complete positivity. From
$L_t$ given on \eqref{eq:lt} of the section~\ref{G} we have its 
inverse matrix
\begin{equation}
L_t^{-1}=\begin{pmatrix}
          r/d & 0 & 0 & (r-1)/d \\
          0  & z_1^*/D & -z_2/D & 0 \\
          0  & -z_2^*/D & z_1/D & 0 \\
          (r-1)/d & 0 & 0 & r/d
         \end{pmatrix}
\end{equation}
where $d=r^2-(1-r)^2=2r-1$ and $D=|z_1|^2-|z_2|^2$. 

If we denote with primes
the functions evaluated on $t+\eps$, that is, $r'=r(t+\eps)$ similarly for
$z_1'$ and $z_2'$, then 
\begin{equation}
L_{t+\eps}=\begin{pmatrix}
          r' & 0 & 0 & 1-r' \\
          0  & z_1' & -z_2' & 0 \\
          0  & z_2'^* & z_1'^* & 0 \\
          1-r' & 0 & 0 & r'
         \end{pmatrix}.
\end{equation}
We found that the super-operator representation corresponding
to $\Lambda_{t+\eps,t}$ is given by
\begin{equation}
L_{t+\eps,t}=L_{t+\eps}L_{t}^{-1}=
\begin{pmatrix}
 q & 0 & 0 & 1-q \\
 0 & Z_1 & Z_2 & 0 \\
 0 & Z_2^* & Z_1^* & 0 \\
 1-q & 0 & 0 & q
\end{pmatrix}
\end{equation}
where $q=(r'-r-1)/d,\; Z_1=(z'_1z_1^*-z'_2z_2^*)/D$ and 
$Z_2=(z'_2z_1-z'_1z_2)/D$. The corresponding Choi matrix turns out of 
reshuffle the above matrix:
\begin{equation}
C_{t+\eps,t}=\begin{pmatrix}
 q & 0 & 0 & Z_1^* \\
 0 & 1-q & Z_2 & 0 \\
 0 & Z_2^* & 1-q^* & 0 \\
 Z_1 & 0 & 0 & q
\end{pmatrix},
\end{equation}
on subsection~\ref{subsec:divi} the divisibility of $\Lambda_{t+\eps,t}$
is explored via the positivity (non negative eigenvalues) of Choi 
matrix $C_{t+\eps,t}$.

%\section{Trace distance under the evolution in a quantum process}
\begin{widetext}
\subsection{Trace distance and contractivity}
\label{app:sec:tracedist}
%On theory of open quantum systems is frequently to use the trace
%distance. 

We have introduced the trace distance and its definition through of 
the eq.~\eqref{eq:sigmaoft}.
%Given two quantum states, $\varrho_1$ and $\varrho_2$, its trace distance 
%is defined as
%\begin{equation}
%\label{eq:app:tracedist}
%T(\varrho_1,\varrho_2)=\frac{1}{2}{\rm tr}(|\varrho_1-\varrho_2|),
%\end{equation}
%where $|A| =\sqrt{A^\dagger A}$.
Furthermore, turns out that the trace distance of an Hermitian matrix
is equal to one half of the sum of absolute values of its 
eigenvalues. 

Given two any states $\varrho_1$ and $\varrho_2$ which evolve under the
quantum channel $\Lambda_t$, its trace distance at the time $t$ can be 
calculate as
\begin{equation}
T[\varrho_1(t),\varrho_2(t)]=\frac{1}{2}{\rm tr}\left(|\Lambda_t[\varrho_1-
\varrho_2]|\right),
\end{equation}
the right side of above equation follows from the linearity of $\Lambda_t$
and due that $\varrho_i(t)=\Lambda_t[\varrho_i]$. Now, if the states 
$\varrho_1$ and $\varrho_2$ are described by the Bloch vectors 
$\vec{a}=(a_x,a_y,a_z)$ and $\vec{b}=(b_x,b_y,b_z)$, respectively. We can
write down
\begin{align}
T[\varrho_1(t),\varrho_2(t)]=\frac{1}{4}{\rm tr}(|\Lambda_t[
c_x(|0\ra\la1|+|1\ra\la0|) +ic_y(|1\ra\la0|-|0\ra\la1|)+
c_z(|0\ra\la0|-|1\ra\la1|)
]|),
\end{align}
where the $c_i$ are the vector components of $\vec{c}=\vec{a}-\vec{b}$.
Using the linearity of $\Lambda_t$  and spherical coordinates for
$\vec{c}=(R\sin\theta\cos\phi,R\sin\theta\sin\phi,R\cos\theta)$ we have
\begin{align}
T[\varrho_1(t),\varrho_2(t)] =\frac{1}{4}{\rm tr}\left(\left|
\mt{R(2r-1)\cos\theta}
{R\rme^{-\rmi\;\phi}\sin\theta(z_1^*+\rme^{\rmi\;2\phi}z_2^*)}
{R\rme^{\rmi\;\phi}\sin\theta(z_1+\rme^{-\rmi\;2\phi}z_2)}
{-R(2r-1)\cos\theta}
\right|\right)
\end{align}
which is an Hermitian matrix therefore
\begin{equation}
T[\varrho_1(t),\varrho_2(t)]=
\frac{R\sqrt{(2r-1)^2\cos^2\theta +
|z_1+z_2\;\rme^{-\rmi\;2\phi}|^2\;\sin^2\theta}}{2}.
\label{eq:app:tdqbgeneral}\end{equation}
\end{widetext}

%\section{\label{apX} Do we need an appendix?} % {{{

\end{appendix} 
% }}}
\bibliographystyle{apsrev} %{{{
\bibliography{./references/JabRef-Deli,./references/moises,./references/carlos}

\begin{thebibliography}{31}
\expandafter\ifx\csname natexlab\endcsname\relax\def\natexlab#1{#1}\fi
\expandafter\ifx\csname bibnamefont\endcsname\relax
  \def\bibnamefont#1{#1}\fi
\expandafter\ifx\csname bibfnamefont\endcsname\relax
  \def\bibfnamefont#1{#1}\fi
\expandafter\ifx\csname citenamefont\endcsname\relax
  \def\citenamefont#1{#1}\fi
\expandafter\ifx\csname url\endcsname\relax
  \def\url#1{\texttt{#1}}\fi
\expandafter\ifx\csname urlprefix\endcsname\relax\def\urlprefix{URL }\fi
\providecommand{\bibinfo}[2]{#2}
\providecommand{\eprint}[2][]{\url{#2}}

\bibitem[{\citenamefont{Landau}(1927)}]{landau}
\bibinfo{author}{\bibfnamefont{L.}~\bibnamefont{Landau}},
  \bibinfo{journal}{Zeitschrift für Physik} \textbf{\bibinfo{volume}{45}},
  \bibinfo{pages}{430} (\bibinfo{year}{1927}).

\bibitem[{\citenamefont{Lindblad}(1976)}]{Lindblad1976}
\bibinfo{author}{\bibfnamefont{G.}~\bibnamefont{Lindblad}},
  \bibinfo{journal}{Communications in Mathematical Physics}
  \textbf{\bibinfo{volume}{48}}, \bibinfo{pages}{119} (\bibinfo{year}{1976}).

\bibitem[{\citenamefont{Gorini et~al.}(1976)\citenamefont{Gorini, Kossakowski,
  and Sudarshan}}]{Gorini1976}
\bibinfo{author}{\bibfnamefont{V.}~\bibnamefont{Gorini}},
  \bibinfo{author}{\bibfnamefont{A.}~\bibnamefont{Kossakowski}},
  \bibnamefont{and} \bibinfo{author}{\bibfnamefont{E.~C.~G.}
  \bibnamefont{Sudarshan}}, \bibinfo{journal}{Journal of Mathematical Physics}
  \textbf{\bibinfo{volume}{17}}, \bibinfo{pages}{821} (\bibinfo{year}{1976}).

\bibitem[{\citenamefont{Prosen}(2008)}]{Prosen2008}
\bibinfo{author}{\bibfnamefont{T.}~\bibnamefont{Prosen}}, \bibinfo{journal}{New
  Journal of Physics} \textbf{\bibinfo{volume}{10}}, \bibinfo{pages}{043026}
  (\bibinfo{year}{2008}).

\bibitem[{\citenamefont{Carmichael}(1999)}]{carmichael1999statistical}
\bibinfo{author}{\bibfnamefont{H.}~\bibnamefont{Carmichael}},
  \emph{\bibinfo{title}{Statistical Methods in Quantum Optics 1: Master
  Equations and Fokker-Planck Equations}}, Physics and Astronomy Online Library
  (\bibinfo{publisher}{Springer}, \bibinfo{year}{1999}), ISBN
  \bibinfo{isbn}{9783540548829}.

\bibitem[{\citenamefont{Breuer et~al.}(2009{\natexlab{a}})\citenamefont{Breuer,
  Laine, and Piilo}}]{PhysRevLett.103.210401}
\bibinfo{author}{\bibfnamefont{H.-P.} \bibnamefont{Breuer}},
  \bibinfo{author}{\bibfnamefont{E.-M.} \bibnamefont{Laine}}, \bibnamefont{and}
  \bibinfo{author}{\bibfnamefont{J.}~\bibnamefont{Piilo}},
  \bibinfo{journal}{Phys. Rev. Lett.} \textbf{\bibinfo{volume}{103}},
  \bibinfo{pages}{210401} (\bibinfo{year}{2009}{\natexlab{a}}).

\bibitem[{\citenamefont{Rivas et~al.}(2010{\natexlab{a}})\citenamefont{Rivas,
  Huelga, and Plenio}}]{Rivas2010}
\bibinfo{author}{\bibfnamefont{{\'A}.}~\bibnamefont{Rivas}},
  \bibinfo{author}{\bibfnamefont{S.}~\bibnamefont{Huelga}}, \bibnamefont{and}
  \bibinfo{author}{\bibfnamefont{M.}~\bibnamefont{Plenio}},
  \bibinfo{journal}{Phys. Rev. Lett.} \textbf{\bibinfo{volume}{105}},
  \bibinfo{pages}{050403} (\bibinfo{year}{2010}{\natexlab{a}}).

\bibitem[{\citenamefont{Breuer}(2012)}]{Breuer2012}
\bibinfo{author}{\bibfnamefont{H.-P.} \bibnamefont{Breuer}},
  \bibinfo{journal}{Journal of Physics B: Atomic, Molecular and Optical
  Physics} \textbf{\bibinfo{volume}{45}}, \bibinfo{pages}{154001}
  (\bibinfo{year}{2012}).

\bibitem[{\citenamefont{Ángel Rivas et~al.}(2014)\citenamefont{Ángel Rivas,
  Huelga, and Plenio}}]{RHP14}
\bibinfo{author}{\bibnamefont{Ángel Rivas}},
  \bibinfo{author}{\bibfnamefont{S.~F.} \bibnamefont{Huelga}},
  \bibnamefont{and} \bibinfo{author}{\bibfnamefont{M.~B.}
  \bibnamefont{Plenio}}, \bibinfo{journal}{Reports on Progress in Physics}
  \textbf{\bibinfo{volume}{77}}, \bibinfo{pages}{094001}
  (\bibinfo{year}{2014}),
  \urlprefix\url{http://stacks.iop.org/0034-4885/77/i=9/a=094001}.

\bibitem[{\citenamefont{Breuer et~al.}(2016)\citenamefont{Breuer, Laine, Piilo,
  and Vacchini}}]{RevModPhys.88.021002}
\bibinfo{author}{\bibfnamefont{H.-P.} \bibnamefont{Breuer}},
  \bibinfo{author}{\bibfnamefont{E.-M.} \bibnamefont{Laine}},
  \bibinfo{author}{\bibfnamefont{J.}~\bibnamefont{Piilo}}, \bibnamefont{and}
  \bibinfo{author}{\bibfnamefont{B.}~\bibnamefont{Vacchini}},
  \bibinfo{journal}{Rev. Mod. Phys.} \textbf{\bibinfo{volume}{88}},
  \bibinfo{pages}{021002} (\bibinfo{year}{2016}),
  \urlprefix\url{https://link.aps.org/doi/10.1103/RevModPhys.88.021002}.

\bibitem[{\citenamefont{de~Vega and Alonso}(2017)}]{RevModPhys.89.015001}
\bibinfo{author}{\bibfnamefont{I.}~\bibnamefont{de~Vega}} \bibnamefont{and}
  \bibinfo{author}{\bibfnamefont{D.}~\bibnamefont{Alonso}},
  \bibinfo{journal}{Rev. Mod. Phys.} \textbf{\bibinfo{volume}{89}},
  \bibinfo{pages}{015001} (\bibinfo{year}{2017}),
  \urlprefix\url{https://link.aps.org/doi/10.1103/RevModPhys.89.015001}.

\bibitem[{\citenamefont{Liu et~al.}(2013)\citenamefont{Liu, Cao, Huang, Li,
  Guo, Laine, Breuer, and Piilo}}]{Liu2013}
\bibinfo{author}{\bibfnamefont{B.-H.} \bibnamefont{Liu}},
  \bibinfo{author}{\bibfnamefont{D.-Y.} \bibnamefont{Cao}},
  \bibinfo{author}{\bibfnamefont{Y.-F.} \bibnamefont{Huang}},
  \bibinfo{author}{\bibfnamefont{C.-F.} \bibnamefont{Li}},
  \bibinfo{author}{\bibfnamefont{G.-C.} \bibnamefont{Guo}},
  \bibinfo{author}{\bibfnamefont{E.-M.} \bibnamefont{Laine}},
  \bibinfo{author}{\bibfnamefont{H.-P.} \bibnamefont{Breuer}},
  \bibnamefont{and} \bibinfo{author}{\bibfnamefont{J.}~\bibnamefont{Piilo}},
  \bibinfo{journal}{Sci. Rep.} \textbf{\bibinfo{volume}{3}},
  \bibinfo{pages}{1781 EP } (\bibinfo{year}{2013}), \bibinfo{note}{article},
  \urlprefix\url{https://doi.org/10.1038/srep01781}.

\bibitem[{\citenamefont{Urrego et~al.}(2018)\citenamefont{Urrego, Fl\'orez,
  Svozil\'{\i}k, Nu\~nez, and Valencia}}]{PhysRevA.98.053862}
\bibinfo{author}{\bibfnamefont{D.~F.} \bibnamefont{Urrego}},
  \bibinfo{author}{\bibfnamefont{J.}~\bibnamefont{Fl\'orez}},
  \bibinfo{author}{\bibfnamefont{J.~c.~v.} \bibnamefont{Svozil\'{\i}k}},
  \bibinfo{author}{\bibfnamefont{M.}~\bibnamefont{Nu\~nez}}, \bibnamefont{and}
  \bibinfo{author}{\bibfnamefont{A.}~\bibnamefont{Valencia}},
  \bibinfo{journal}{Phys. Rev. A} \textbf{\bibinfo{volume}{98}},
  \bibinfo{pages}{053862} (\bibinfo{year}{2018}),
  \urlprefix\url{https://link.aps.org/doi/10.1103/PhysRevA.98.053862}.

\bibitem[{\citenamefont{Carrera et~al.}(2014)\citenamefont{Carrera, Gorin, and
  Seligman}}]{CGS14}
\bibinfo{author}{\bibfnamefont{M.}~\bibnamefont{Carrera}},
  \bibinfo{author}{\bibfnamefont{T.}~\bibnamefont{Gorin}}, \bibnamefont{and}
  \bibinfo{author}{\bibfnamefont{T.~H.} \bibnamefont{Seligman}},
  \bibinfo{journal}{Phys. Rev. A} \textbf{\bibinfo{volume}{90}},
  \bibinfo{pages}{022107} (\bibinfo{year}{2014}),
  \urlprefix\url{http://link.aps.org/doi/10.1103/PhysRevA.90.022107}.

\bibitem[{\citenamefont{Mehta}(2004)}]{Meh2004}
\bibinfo{author}{\bibfnamefont{M.~L.} \bibnamefont{Mehta}},
  \emph{\bibinfo{title}{random matrices and the statistical theory of energy
  levels, 3rd Edition}} (\bibinfo{publisher}{Academic Press},
  \bibinfo{address}{New York}, \bibinfo{year}{2004}).

\bibitem[{\citenamefont{Gardiner et~al.}(1997)\citenamefont{Gardiner, Cirac,
  and Zoller}}]{GCZ97}
\bibinfo{author}{\bibfnamefont{S.~A.} \bibnamefont{Gardiner}},
  \bibinfo{author}{\bibfnamefont{J.~I.} \bibnamefont{Cirac}}, \bibnamefont{and}
  \bibinfo{author}{\bibfnamefont{P.}~\bibnamefont{Zoller}},
  \bibinfo{journal}{Physical Review Letters} \textbf{\bibinfo{volume}{79}},
  \bibinfo{pages}{4790} (\bibinfo{year}{1997}).

\bibitem[{\citenamefont{Gorin et~al.}(2004)\citenamefont{Gorin, Prosen,
  Seligman, and Strunz}}]{GPSS04}
\bibinfo{author}{\bibfnamefont{T.}~\bibnamefont{Gorin}},
  \bibinfo{author}{\bibfnamefont{T.}~\bibnamefont{Prosen}},
  \bibinfo{author}{\bibfnamefont{T.~H.} \bibnamefont{Seligman}},
  \bibnamefont{and} \bibinfo{author}{\bibfnamefont{W.~T.}
  \bibnamefont{Strunz}}, \bibinfo{journal}{Physical Review A: Atomic,
  Molecular, and Optical Physics} \textbf{\bibinfo{volume}{70}},
  \bibinfo{pages}{042105:1} (\bibinfo{year}{2004}).

\bibitem[{\citenamefont{Choi}(1975)}]{Choi1975285}
\bibinfo{author}{\bibfnamefont{M.-D.} \bibnamefont{Choi}},
  \bibinfo{journal}{Linear Algebra and its Applications}
  \textbf{\bibinfo{volume}{10}}, \bibinfo{pages}{285 } (\bibinfo{year}{1975}),
  ISSN \bibinfo{issn}{0024-3795},
  \urlprefix\url{http://www.sciencedirect.com/science/article/pii/0024379575900750}.

\bibitem[{\citenamefont{Ziman and Heinosaari}(2012)}]{zimanbook2012}
\bibinfo{author}{\bibfnamefont{M.}~\bibnamefont{Ziman}} \bibnamefont{and}
  \bibinfo{author}{\bibfnamefont{T.}~\bibnamefont{Heinosaari}},
  \emph{\bibinfo{title}{The Mathematical Language of Quantum Theory}}
  (\bibinfo{publisher}{Cambridge University Press}, \bibinfo{year}{2012}).

\bibitem[{\citenamefont{Rivas et~al.}(2010{\natexlab{b}})\citenamefont{Rivas,
  Huelga, and Plenio}}]{RiHuPl10}
\bibinfo{author}{\bibfnamefont{A.}~\bibnamefont{Rivas}},
  \bibinfo{author}{\bibfnamefont{S.~F.} \bibnamefont{Huelga}},
  \bibnamefont{and} \bibinfo{author}{\bibfnamefont{M.~B.}
  \bibnamefont{Plenio}}, \bibinfo{journal}{Phys. Rev. Lett.}
  \textbf{\bibinfo{volume}{105}}, \bibinfo{pages}{050403}
  (\bibinfo{year}{2010}{\natexlab{b}}),
  \urlprefix\url{http://link.aps.org/doi/10.1103/PhysRevLett.105.050403}.

\bibitem[{\citenamefont{van Kampen}(2007)}]{vKampen07}
\bibinfo{author}{\bibfnamefont{N.~G.} \bibnamefont{van Kampen}},
  \emph{\bibinfo{title}{Stochastic Processes in Physics and Chemistry}}
  (\bibinfo{publisher}{North-Holland, Elsevier}, \bibinfo{year}{2007}).

\bibitem[{\citenamefont{Breuer et~al.}(2009{\natexlab{b}})\citenamefont{Breuer,
  Laine, and Piilo}}]{BrLaPi09}
\bibinfo{author}{\bibfnamefont{H.-P.} \bibnamefont{Breuer}},
  \bibinfo{author}{\bibfnamefont{E.-M.} \bibnamefont{Laine}}, \bibnamefont{and}
  \bibinfo{author}{\bibfnamefont{J.}~\bibnamefont{Piilo}},
  \bibinfo{journal}{Phys. Rev. Lett.} \textbf{\bibinfo{volume}{103}},
  \bibinfo{pages}{210401} (\bibinfo{year}{2009}{\natexlab{b}}),
  \urlprefix\url{http://link.aps.org/doi/10.1103/PhysRevLett.103.210401}.

\bibitem[{\citenamefont{Nielsen and Chuang}(2000)}]{NieChu00}
\bibinfo{author}{\bibfnamefont{M.~A.} \bibnamefont{Nielsen}} \bibnamefont{and}
  \bibinfo{author}{\bibfnamefont{I.~L.} \bibnamefont{Chuang}},
  \emph{\bibinfo{title}{quantum computation and quantum information}}
  (\bibinfo{publisher}{Cambridge University Press},
  \bibinfo{address}{Cambridge}, \bibinfo{year}{2000}).

\bibitem[{\citenamefont{Wi\ss{}mann et~al.}(2012)\citenamefont{Wi\ss{}mann,
  Karlsson, Laine, Piilo, and Breuer}}]{Wis12}
\bibinfo{author}{\bibfnamefont{S.}~\bibnamefont{Wi\ss{}mann}},
  \bibinfo{author}{\bibfnamefont{A.}~\bibnamefont{Karlsson}},
  \bibinfo{author}{\bibfnamefont{E.-M.} \bibnamefont{Laine}},
  \bibinfo{author}{\bibfnamefont{J.}~\bibnamefont{Piilo}}, \bibnamefont{and}
  \bibinfo{author}{\bibfnamefont{H.-P.} \bibnamefont{Breuer}},
  \bibinfo{journal}{Phys. Rev. A} \textbf{\bibinfo{volume}{86}},
  \bibinfo{pages}{062108} (\bibinfo{year}{2012}),
  \urlprefix\url{http://link.aps.org/doi/10.1103/PhysRevA.86.062108}.

\bibitem[{\citenamefont{Chru\ifmmode \acute{s}\else
  \'{s}\fi{}ci\ifmmode~\acute{n}\else \'{n}\fi{}ski
  et~al.}(2011)\citenamefont{Chru\ifmmode \acute{s}\else
  \'{s}\fi{}ci\ifmmode~\acute{n}\else \'{n}\fi{}ski, Kossakowski, and
  Rivas}}]{Chr11}
\bibinfo{author}{\bibfnamefont{D.}~\bibnamefont{Chru\ifmmode \acute{s}\else
  \'{s}\fi{}ci\ifmmode~\acute{n}\else \'{n}\fi{}ski}},
  \bibinfo{author}{\bibfnamefont{A.}~\bibnamefont{Kossakowski}},
  \bibnamefont{and} \bibinfo{author}{\bibfnamefont{A.}~\bibnamefont{Rivas}},
  \bibinfo{journal}{Phys. Rev. A} \textbf{\bibinfo{volume}{83}},
  \bibinfo{pages}{052128} (\bibinfo{year}{2011}),
  \urlprefix\url{http://link.aps.org/doi/10.1103/PhysRevA.83.052128}.

\bibitem[{\citenamefont{Cabrera et~al.}(2019)\citenamefont{Cabrera, Davalos,
  and Gorin}}]{MoDaGo19}
\bibinfo{author}{\bibfnamefont{G.~M.} \bibnamefont{Cabrera}},
  \bibinfo{author}{\bibfnamefont{D.}~\bibnamefont{Davalos}}, \bibnamefont{and}
  \bibinfo{author}{\bibfnamefont{T.}~\bibnamefont{Gorin}},
  \bibinfo{journal}{Phys. Lett. A (in press)} \bibinfo{eid}{arXiv:1902.06829}
  (\bibinfo{year}{2019}).

\bibitem[{\citenamefont{Pineda et~al.}(2016)\citenamefont{Pineda, Gorin,
  Davalos, Wisniacki, and Garc\'{\i}a-Mata}}]{PGDWG16}
\bibinfo{author}{\bibfnamefont{C.}~\bibnamefont{Pineda}},
  \bibinfo{author}{\bibfnamefont{T.}~\bibnamefont{Gorin}},
  \bibinfo{author}{\bibfnamefont{D.}~\bibnamefont{Davalos}},
  \bibinfo{author}{\bibfnamefont{D.~A.} \bibnamefont{Wisniacki}},
  \bibnamefont{and}
  \bibinfo{author}{\bibfnamefont{I.}~\bibnamefont{Garc\'{\i}a-Mata}},
  \bibinfo{journal}{Phys. Rev. A} \textbf{\bibinfo{volume}{93}},
  \bibinfo{pages}{022117} (\bibinfo{year}{2016}),
  \urlprefix\url{http://link.aps.org/doi/10.1103/PhysRevA.93.022117}.

\bibitem[{\citenamefont{Breuer and Petruccione}(2002)}]{BrePet02}
\bibinfo{author}{\bibfnamefont{H.-P.} \bibnamefont{Breuer}} \bibnamefont{and}
  \bibinfo{author}{\bibfnamefont{F.}~\bibnamefont{Petruccione}},
  \emph{\bibinfo{title}{The Theory of open quantum systems}}
  (\bibinfo{publisher}{Oxford University Press}, \bibinfo{year}{2002}).

\bibitem[{\citenamefont{Merzbacher}(1998)}]{merzbacher1998quantum}
\bibinfo{author}{\bibfnamefont{E.}~\bibnamefont{Merzbacher}},
  \emph{\bibinfo{title}{Quantum Mechanics}} (\bibinfo{publisher}{Wiley},
  \bibinfo{year}{1998}), ISBN \bibinfo{isbn}{9780471887027},
  \urlprefix\url{https://books.google.com.mx/books?id=6Ja\_QgAACAAJ}.

\bibitem[{\citenamefont{Roa et~al.}(2014)\citenamefont{Roa, Mu\~noz, and
  Gr\"uning}}]{RoMuGr14}
\bibinfo{author}{\bibfnamefont{L.}~\bibnamefont{Roa}},
  \bibinfo{author}{\bibfnamefont{A.}~\bibnamefont{Mu\~noz}}, \bibnamefont{and}
  \bibinfo{author}{\bibfnamefont{G.}~\bibnamefont{Gr\"uning}},
  \bibinfo{journal}{Phys. Rev. A} \textbf{\bibinfo{volume}{89}},
  \bibinfo{pages}{064301} (\bibinfo{year}{2014}),
  \urlprefix\url{https://link.aps.org/doi/10.1103/PhysRevA.89.064301}.

\bibitem[{\citenamefont{Bengtsson and Zyczkowski}(2006)}]{BenZyc06}
\bibinfo{author}{\bibfnamefont{I.}~\bibnamefont{Bengtsson}} \bibnamefont{and}
  \bibinfo{author}{\bibfnamefont{K.}~\bibnamefont{Zyczkowski}},
  \emph{\bibinfo{title}{Geometry of Quantum States: An Introduction to Quantum
  Entanglement}} (\bibinfo{publisher}{Cambridge University Press},
  \bibinfo{year}{2006}).

\end{thebibliography}
% }}}
\end{document}